\documentclass[12pt]{iopart}
\usepackage{iopams}
\usepackage{graphicx}
\usepackage{epstopdf}

\begin{document}
\title{Non-local response in a lattice gas under a shear drive}

\author{Tridib Sadhu}
\address{Institut de Physique Th\'{e}orique, CEA/Saclay, F-91191 Gif-sur-Yvette Cedex, France.}
\author{Satya N. Majumdar}
\address{Univ. Paris-Sud, CNRS, LPTMS, UMR 8626, Orsay F-01405,
France.}
\author{David Mukamel}
\address{Physics of Complex Systems, Weizmann Institute of Science, Rehovot 76100, Israel.}
\date{}

\begin{abstract}
	In equilibrium, the effect of a spatially localized perturbation is
	typically confined around the perturbed region. Quite contrary to this,
	in a non-equilibrium stationary state often the entire system is
	affected. This appears to be a generic feature of non-equilibrium.
	We study such non-local response in the stationary state of a lattice gas 		with a shear drive
	at the boundary which keeps the system out of equilibrium. We
	show that a perturbation in the form of a localized blockage at the
	boundary, induces algebraically decaying density and current profile.
	In two examples, non-interacting particles and particles with simple exclusion,
	we analytically derive the power-law tail of the profiles.
\end{abstract}
\vspace{2pc}

\submitto{\JPA}

\maketitle
\section{Introduction}
Systems far from equilibrium often display some novel and unexpected features
that are in striking contrast to equilibrium \cite{CHOU,EVANS,HENKEL}.
One of the intriguing features of non-equilibrium stationary state is
the presence of long-range correlation at generic parameter values
\cite{GRINSTEIN_REVIEW}.
It is well known that non-equilibrium stationary state of systems evolving under a dynamics
which conserves some variables (such as energy, momentum or density)
display slow decay of correlations, even when the dynamics
is local \cite{ZIA}. This is observed in several studies of
driven diffusive systems \cite{SPOHN,KLS1,KLS2,ZHANG,GARRIDO,GRINSTEIN,RUBI,SASA,SMM3}.

A consequence of the long-range correlation is that a local
perturbation can lead to global changes in the one-point function, like density profile. Such non-local response has
been demonstrated earlier in a system of diffusive particles \cite{Maes} and also in a lattice-gas with hard-core repulsion (simple exclusion) \cite{SMM3,SMM}. In this paper we investigate a scenario different from those studied earlier: the non-equilibrium stationary state of a lattice-gas under a shear drive along the boundary. Our purpose is to analyze the effect of a localized perturbation on this stationary state.

An alternate motivation comes from studies in fluid medium, where a shear flow localized at the boundary often leads to non-local changes in the liquid structure \cite{ONUKI_KAWASAKI,ONUKI,BINDER,BICKEL}. Here the long-range effect is related to the slow modes of the capillary wave which originates essentially from momentum conservation \cite{BICKEL}. A natural question to ask is, what happens when there is no inertia, and hence no convective current which can carry the
information of the boundary flow to the bulk. Does the long-range effect survive? Such momentum non-conserving dynamics are important in describing processes within biological cells or mobility of individual cells --- the world of low Reynolds number.

Our study is based on a two-dimensional lattice gas with cylindrical boundary condition. The shear drive is at the bottom lane which biases the motion of particles in one direction along the lane (see \Fref{fig:fig1}). In the bulk, there is no drive and the particles move symmetrically across any
bond. For our theoretical analysis we consider two cases: one, where the particles are independent of each other, and the second, where the particles interact with simple exclusion such that no two particles can occupy the same site at the same time. In the latter case, the model is essentially a symmetric simple exclusion process (SSEP) coupled to a totally asymmetric simple exclusion process (TASEP) at the boundary. The particles don't have momentum and the only conserved quantity is the total particle number.

In both examples considered, the boundary shear drive, in itself, does not induce any current in the bulk and the
average density profile remains the same as in the absence of drive. However, if a defect is introduced on a single bond at the boundary lane which hinders the
motion across the bond, particles on the boundary lane go around
the blockage by hopping to neighboring lanes. This way the boundary current is transmitted to the
bulk. The defect at the boundary lane can be considered as a local perturbation on the non-equilibrium stationary state with shear drive. What is surprising is that the induced current is not just confined
at the neighborhood of the boundary, it extends even far from it. In fact,
the amplitude of the current decays algebraically with the distance from the boundary. The density profile of the particles is also affected globally,
where the difference in density from a flat profile has a power-law tail.

For the example of independent particles, we rigorously derive the current and the density profile. For the exclusion interactions, an exact analysis becomes involved as there is hierarchical dependence of correlations. We circumvent this hierarchy by using a mean-field approximation. However a derivation of the profiles is still quite difficult as it requires solving an infinite set of coupled non-linear equations. We have thus extracted the large distance profile using a perturbation expansion in the strength of the blockage. Our analysis reveals that in both types of interaction, the density decays as $1/r$ with the
distance $r$ from the blockage, whereas the current decays as $1/r^{2}$. We argue from a hydrodynamic description that the power-law decay is the same even for other short-range interactions among particles.

Our result has direct relevance to several other systems. For example, the model is closely related to the directed motion of molecular motors along a lane made of cytoskeletal filament \cite{HOWARD}. The motors are actively driven in one direction when they are attached to a cytoskeletal filament. Due to the thermal
fluctuations, the motors usually unbind after a certain time from the filament. While unbound, the
motors diffuse freely in the surrounding fluid until they eventually
re-attach to the filament. Lipowsky and co-workers have modelled this
as a TASEP on a single lane coupled to a SSEP on the surrounding multiple lanes
\cite{KLUMPP}. Our result shows that a defect on the cytoskeletal
filament generates current around the defect with a slowly decaying amplitude. This should be accessible to experiments in biological transport.

Our model is also an example of coupled multi-lane exclusion process. In particular, the model is a multi-lane generalization of a well known model of one-dimensional TASEP with a slow
bond, first studied by Janowsky and Lebowitz in 1992 \cite{JANOWSKY1}.
An exact characterization of the stationary properties of the J-L model is still elusive \cite{JANOWSKY1,JANOWSKY2,SCHUTZ,TIMO}.
There have been several other studies on the multi-lane transport \cite{SALERNO,EZAKI,POTIGUAR}.
They were mostly introduced to model biological transport \cite{HOWARD,KLUMPP}, transport of spins in quantum systems
\cite{REICHENBACH}, macroscopic clustering phenomena \cite{KORNISS}, and also in vehicular
traffic \cite{HELBING,CHOWDHURY,LEE,KANAI}. In the case of two lanes with opposite bias, competing currents on the coupled lanes often produce complex dynamical behavior and rich phase diagram
\cite{PRONINA1,PRONINA2,HARRIS,REICHENBACH,JIANG,SCHIFFMANN,MELBINGER,POPKOV,LEE,MITSUDO,TSEKOURAS,DICKMAN}.
Particularly relevant to our model is the work in \cite{TSEKOURAS,YADAV} where a single lane asymmetric exclusion
process is coupled to a single lane symmetric exclusion process.
The idea of surface drive in a multi-lane lattice-gas has also been
used in kinetic Ising model for simulating magnetic friction
\cite{KADAU,HUCHT,HILHORST}.

We present the results of our work in the following order. In \sref{sec:model}, we
introduce the model for the two cases: independent particles and particles with simple exclusion.
In \sref{sec:ni}, we present a detailed analysis of the independent particles
case. In \sref{sec:excl}, we
discuss the model with exclusion interactions. The rate equation for the density is solved in a perturbative expansion. For simplicity we first
present the calculation of the leading term in the expansion. The higher order terms
can be determined recursively and are discussed in the Appendix. We show that inclusion of the
higher order terms does not change the power-law tail of the density profile.
In \sref{sec:electro}, we argue for the algebraic profile using an electrostatic
analogy. Finally, in \sref{sec:summary}, we summarize by discussing the robustness of the
power-law tail in lattice-gas with beyond simple exclusion interaction.

\section{The model \label{sec:model}}
\begin{figure}
\begin{center}
\includegraphics[width=10.0cm]{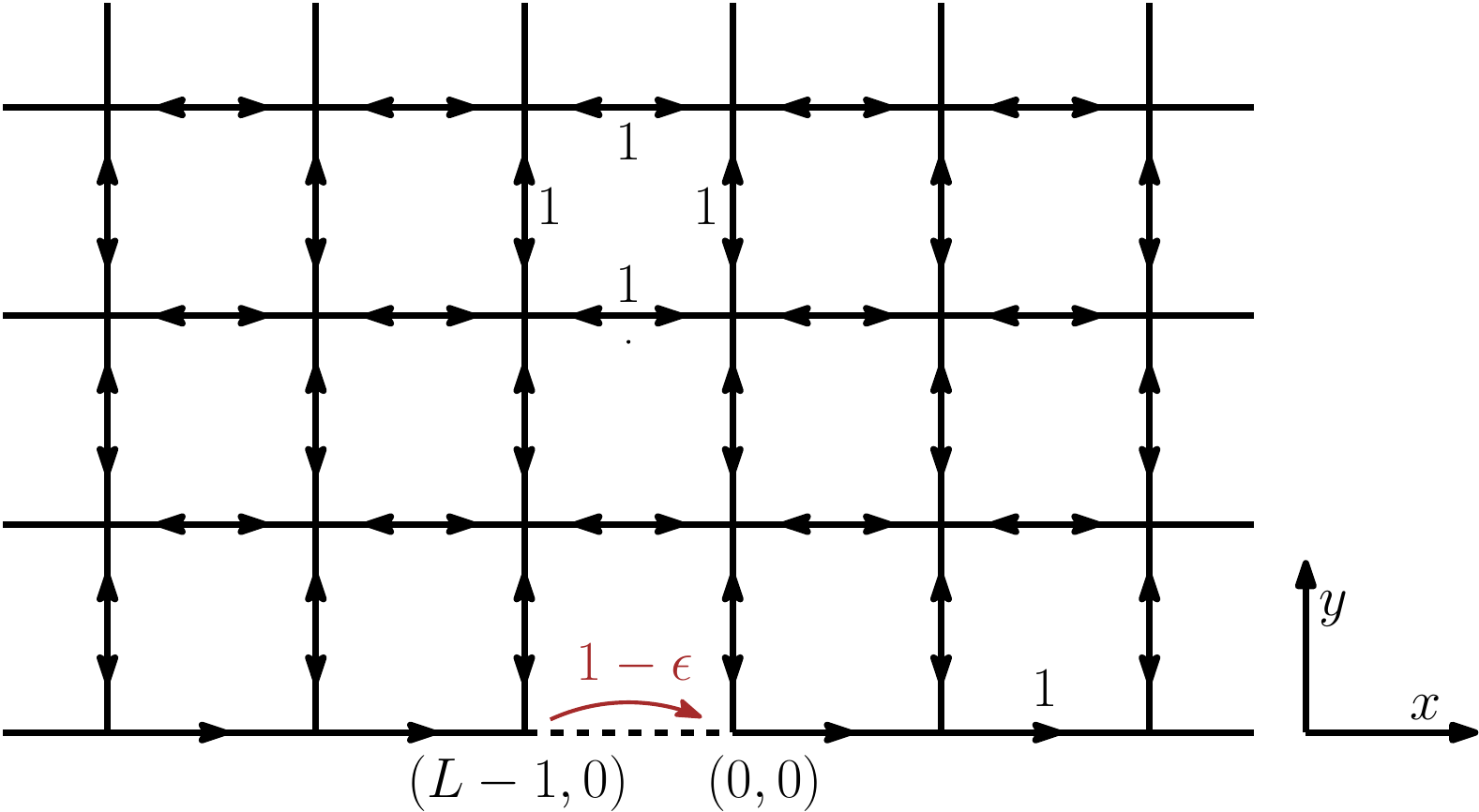}
\caption{The model on a square lattice with periodic boundary
condition along the $x$-axis, and reflecting boundary condition along
the $y$-axis. The jump rate of particles in the bottom lane is totally asymmetric, with anti-clockwise jumps forbidden. The slow bond is between sites $\left( L-1,0 \right)$ and $\left(
0,0 \right)$.}
\label{fig:fig1}
\end{center}
\end{figure}

We consider a lattice gas on a $[0,L-1]\times[0,M-1]$ square lattice with
cylindrical boundary condition, \textit{i.e.}, with periodicity along the
$x$ direction and reflecting boundary in the $y$ direction (see
\Fref{fig:fig1}). Particles in the bulk diffuse symmetrically by
jumping to the neighboring sites. We consider two cases: (a) independent
particles and (b) particles with hard core repulsion (simple exclusion). In the latter case, a jump is allowed only when the destination site is empty.
At the boundary lane $y=0$, there is a shear drive which forbids anti-clockwise jumps along the lane. In addition, there is a defect bond on this lane where the jump rate is slower than the rest of the lattice.
The time scale is set by taking the jump rate across this slow bond
as $1-\epsilon$, and across the rest of the bonds as $1$.

In summary, the jump rates are the following:
\begin{enumerate}
\item
Across the bond between sites $(L-1,0)$ and $(0,0)$, particles jump
only clockwise, with rate $1-\epsilon$, where $0\le \epsilon\le 1$.
\item
At the rest of the bonds in the $y=0$ lane, the clockwise jump rate is
$1$. Anti-clockwise hops are forbidden.
\item
For any other bond in the $y>0$ lanes the jump is symmetric with rate $1$.
\item
At the top most lane $y=M-1$ there are only three types of jumps
allowed, two within the lane and one to the neighbor lane $y=M-2$. All
of them are with rate $1$.
\end{enumerate}
The total number of particles is conserved at any time.

\section{Independent particles \label{sec:ni}}
In absence of the boundary drive, \textit{i.e.}, when the jump rates everywhere are symmetric, the dynamics satisfy detailed balance, and the stationary state is in equilibrium. It is easy to see that the average
density of particles per site is uniform across the lattice, and equal
to $\rho=N/LM$, where $N$ is the total number of particles.

\begin{figure}
\begin{center}
\includegraphics[width=8.0cm]{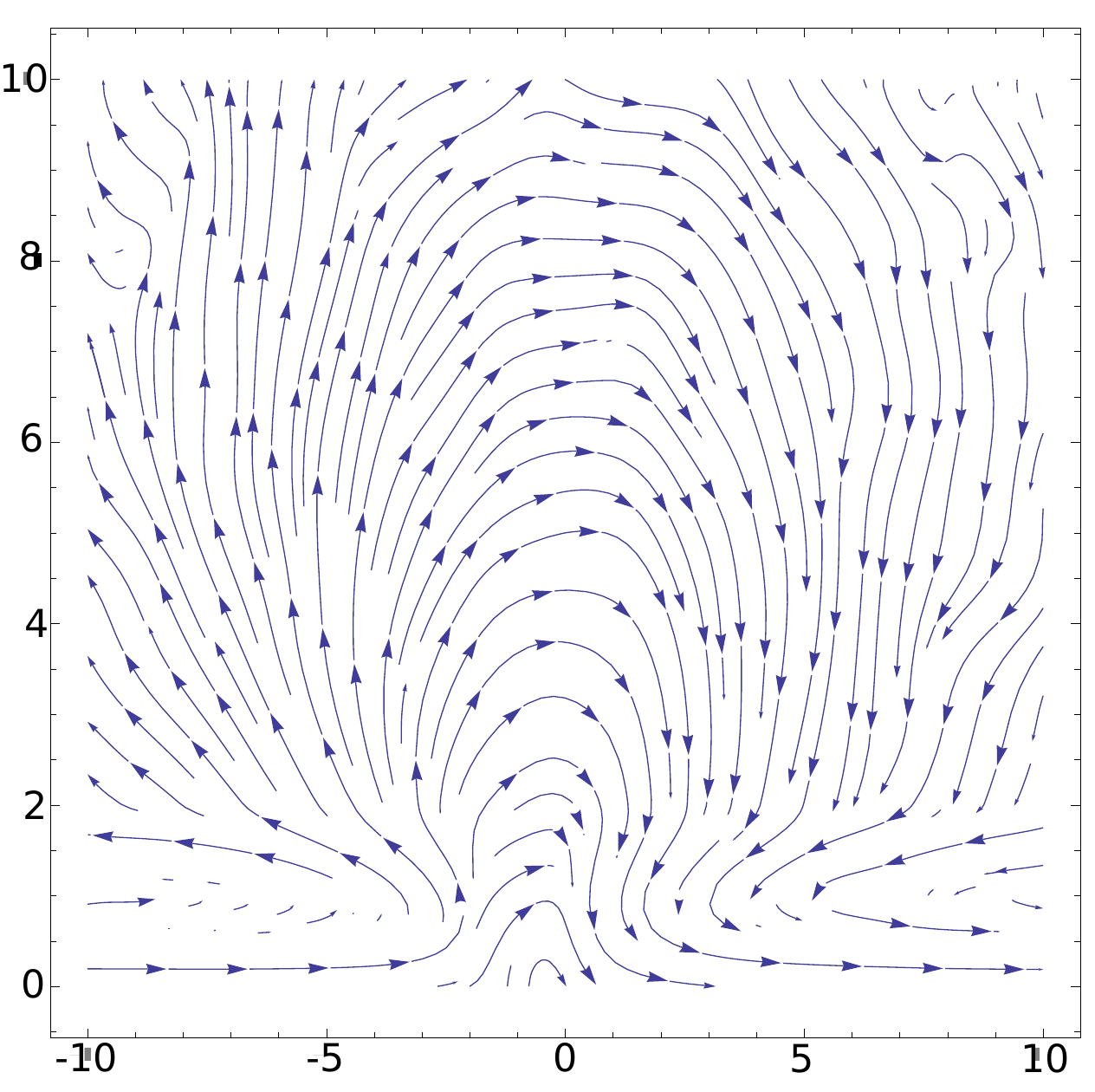}
\caption{A streamline plot of the stationary state current of
non-interacting particles generated in a Monte Carlo simulation on a
$100\times100$ square lattice. The figure is centered around the
broken bond placed between the sites $\left( -1,0 \right)$ and $\left(
0,0 \right)$.}
\label{fig:cur}
\end{center}
\end{figure}

When the boundary drive is switched on, the asymmetric jump rate drives a particle current along the boundary lane. Because of the slow jump rate across the defect bond, there is a density gradient of particles around it. This induces a diffusive current in the neighboring lanes. A stream line
plot of the currents generated using Monte Carlo simulation is shown
in \fref{fig:cur}. We have not shown the current on the driven lane as it
overshadows the diffusive current on the same scale.

\begin{figure}
\begin{center}
\includegraphics[width=10.0cm]{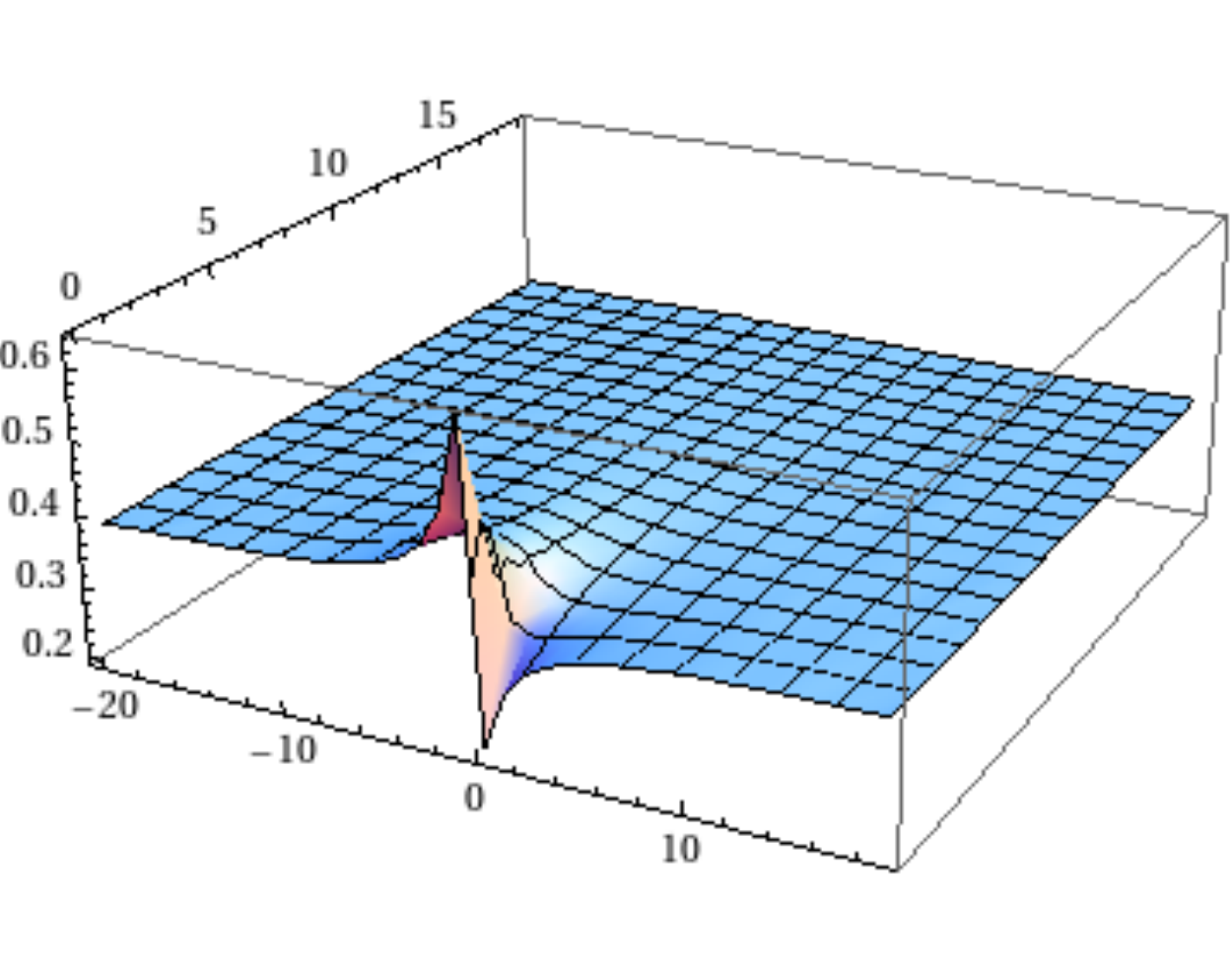}
\caption{The stationary state density profile of the independent
particles diffusing on a $600\times 1200$ square lattice. The slow bond is between sites $\left(-1,0 \right)$ and $\left( 0,0 \right)$. The global average density is $0.4$.}
\label{fig:density}
\end{center}
\end{figure}

In the stationary state, the density is maximum at the site
$(L-1,0)$, and is minimum at $(0,0)$. Away from the blockage, the
density approaches the global average value $\rho$.
An example of the stationary density profile generated by Monte Carlo
simulation is shown in \fref{fig:density}. In the following, we shall show
that the density per site, as well as the induced current
decay algebraically with the distance from the defect bond.

For simplicity we take $M\rightarrow\infty$, and
consider a semi-infinite lattice. Let us define the density $\phi_{t}(x,y)$ as the ensemble averaged number of particles at site $(x,y)$ at time
$t$. Following the dynamics given in \sref{sec:model}, it is easy to
write down the time evolution of $\phi_{t}(x,y)$. Using the rules
$(i)$ and $(ii)$, for $y=0$ we get
\begin{equation}
\frac{\partial \phi_{t}(x,0)}{\partial
t}=\phi_{t}(x-1,0)-2\phi_{t}(x,0)+\phi_{t}(x,1)-\epsilon \phi_{t}(L-1,0)\left[
\delta_{x,0}-\delta_{x,L-1} \right].
\end{equation}
The two Kronecker delta functions are due to the slow bond between
the sites $\left( L-1,0 \right)$ and $\left( 0,0 \right)$.
The periodic boundary condition is imposed by defining $(-1,y)\equiv
(L-1,y)$ and $(L,y)\equiv (0,y)$. Similarly, using the rule $(iii)$ for
$y>0$ yields
\begin{equation}
\frac{\partial \phi_{t}(x,y)}{\partial
t}=\phi_{t}(x-1,y)+\phi_{t}(x+1,y)+\phi_{t}(x,y+1)+\phi_{t}(x,y-1)-4\phi_{t}(x,y).
\end{equation}

In the stationary state where the time derivative vanishes, the equations yield
\begin{eqnarray}
\fl \qquad \phi(x-1,0)+\phi(x,1)-2\phi(x,0)=\epsilon\phi(L-1,0)\left[\delta_{x,0}-\delta_{x,L-1}\right]
&\textrm{ for \ensuremath{y=0}}\label{eq:masternon1}\\
\fl \qquad \Delta\phi(x,y)= 0&\textrm{ for \ensuremath{y>0},}
\label{eq:masternon2}
\end{eqnarray}
where we dropped the time index, and also defined discrete Laplacian $\Delta$ by,
\begin{equation}
\Delta\phi(x,y)=\phi(x-1,y)+\phi(x+1,y)+\phi(x,y+1)+\phi(x,y-1)-4\phi(x,y).
\label{eq:dl}
\end{equation}
The solution of \eref{eq:masternon1} and  \eref{eq:masternon2}, with
the boundary condition that far from the driven lane the stationary
density $\phi(x,y)$ approaches the global average $\rho$, determines the
stationary profile.

An interesting feature to note is that the average density for each
lane $L^{-1}\sum_{x}\phi\left( x,y \right)$ is the same as that
in absence of the drive. This can be verified easily from the stationary state equations above. This property will be used later in the derivation.

Equations \eref{eq:masternon1}-\eref{eq:masternon2} are a set of coupled linear equations,
and their solution can be determined exactly. To begin with, we consider
$\epsilon \phi\left( L-1,0\right)=Q$ as a free parameter.
Its value will be determined self-consistently at the end of the calculation.

Due to the cylindrical boundary conditions, the stationary solution
is periodic in the $x$ coordinate, so that,
$\phi(x+L,y)=\phi(x,y)$. Then, the normal modes of the density profile are the Fourier transform
\begin{equation}
g(n,y)=\frac{1}{L}\sum_{x=0}^{L-1}e^{-i\omega_{n}x}\phi(x,y),
\label{eq:fourier}
\end{equation}
with $\omega_{n}=2\pi n/L$ and $n=0,1,\cdots,(L-1)$. The stationary state
equations \eref{eq:masternon1}-\eref{eq:masternon2} in terms of these normal modes yield
\begin{eqnarray}
g(n,1)+\left[e^{-i\omega_{n}}-2\right]g(n,0)=\frac{Q}{L}\left[1-e^{i\omega_{n}}\right]&\qquad \textrm{ for \ensuremath{y=0},}\label{eq:gboundary1}\\
g(n,y+1)+g(n,y-1)=2\left[2-\cos\omega_{n}\right]g(n,y) &\qquad \textrm{ for \ensuremath{y>0}.}
\label{eq:gboundary2}
\end{eqnarray}

As mentioned earlier, the average density per lane is $\rho$ which yields $g(0,y)=\rho$, for all $y$.
Then the stationary state profile in terms of these normal modes is
\begin{equation}
\phi\left( x,y \right)=\rho+\sum_{n=1}^{L-1}g\left(n, y \right)e^{i\omega_{n}x}.
\label{eq:invfourier}
\end{equation}

The Fourier amplitude $g(n,y)$, for $n\ge 1$, can be determined
iteratively in terms of $g(n,0)$, using the recurrence
relation \eref{eq:gboundary1} and \eref{eq:gboundary2}.
To perform this calculation systematically, let us define the generating function
\begin{equation}
G(n,z)=\sum_{y=1}^{\infty}g\left(n, y \right)z^{y}, \qquad \textrm{for all
$n>0$}.
\end{equation}
Note that $g(n,0)$ has been excluded from the definition. Using
\eref{eq:gboundary2} it is easy to show that
\begin{equation}
G(n,z)=\frac{z g(n,1)-z^{2}g(n,0)}{(z-z_{-})(z+z_{+})},
\label{gz}
\end{equation}
where
\begin{equation}
z_{\pm}(\omega_{n})=2-\cos\omega_{n}\pm\sqrt{\left(2-\cos\omega_{n}\right)^{2}-1}.
\label{eq:zpm}
\end{equation}

Note that $g(n,0)$ and $g(n,1)$ are so far unknown variables
related by \eref{eq:gboundary1}. A second independent relation can be
found following a pole-cancelling mechanism
\cite{rajesh1,rajesh2} which uses the convergence of the generating
function. From equation \eref{eq:fourier} it is easy to see that the amplitude
of the normal modes are bounded $\vert g(n,y) \vert\le
L^{-1}\sum_{x}\phi(x,y)=\rho$. This implies that, the
generating function $G(n,z)$ for all $n$ converges at any value of $\vert z\vert <1$.
On the other hand, it easy to show that $z_{-}<1$ whereas $z_{+}>1$. Then, to be consistent, $z=z_{-}$ must not be a pole of the
generating function $G(n,z)$, implying that the numerator in equation
\eref{gz} must vanish at $z=z_{-}$. In other words,
\begin{equation}
g(n,1)=z_{-}~g(n,0) \textrm{ for all } n > 0.
\end{equation}

This, together with \eref{eq:gboundary1} determines $g(n,0)$ and $g(n,1)$
in terms of $Q$. The generating function can thus be expressed as
\begin{equation}
G(n,z)=\frac{Q}{L}\frac{1-e^{i\omega_{n}}}{(2-e^{-i\omega_n}-z_{-})}\frac{z}{(z-z_{+})}
\qquad \textrm{ for all \ensuremath{n>0}.}
\label{eq:peq}
\end{equation}

The amplitudes $g(n,y)$ for all $y$, can be extracted from this expression by
expanding \eref{eq:peq} in a Taylor series around $z=0$ and comparing it with the definition of $G(n,z)$.
\begin{equation}
G(n,z)=\sum_{y=1}^{\infty}\left[\frac{Q}{L}\frac{1-e^{i\omega_{n}}}{z_{-}-2+e^{-i\omega_{n}}}\frac{1}{z_{+}^{y}}\right]
z^{y}.
\end{equation}
By definition, the term inside the parenthesis is $g(n,y)$, for $n>0$, which when combined with \eref{eq:invfourier}, yields the stationary density profile,
\begin{equation}
\phi(x,y)=\rho+\frac{Q}{L}\sum_{n=1}^{L-1}\gamma(\omega_{n})\frac{e^{i\omega x}}{z_{+}^{y}},
\label{eq:nonintdensity}
\end{equation}
where we defined
\begin{equation}
\gamma(\omega_{n})=\frac{e^{i\omega_{n}}-1}{2-z_{-}-e^{-i\omega_{n}}}.
\label{eq:gamma}
\end{equation}

The only remaining quantity to be determined is $Q$ which appears as an overall normalization constant of
the density difference. This can be evaluated using the
self-consistency condition $Q=\epsilon\phi\left( L-1,0 \right)$.
This yields
\begin{equation}
Q=\epsilon\rho+\epsilon\frac{Q}{L}\sum_{n=1}^{L-1}\gamma(\omega_{n})e^{-i\omega_{n}}.
\label{eq:selfconsistency}
\end{equation}

The analysis is simpler in the $L\rightarrow
\infty$ limit, where $\omega_{n}\equiv\omega$ can be considered as a continuous
variable, and we replace the summation by integration. Then,
\begin{equation}
Q=\epsilon\rho+\epsilon\frac{Q}{2\pi}\int_{0}^{2\pi}d\omega~ \gamma(\omega)e^{-i\omega_{n}}.
\label{eq:selfconsistencycont}
\end{equation}
Performing the integration (see \ref{app:I1}) and simplifying, we get
\begin{equation}
Q=\frac{2\pi\epsilon \rho}{\left[2\pi-\epsilon\left( \pi+1 \right)\right]}.
\label{eq:Q}
\end{equation}
The complete solution for the density profile is then given by
\begin{equation}
\phi(x,y)-\rho=\frac{\rho\epsilon}{2\pi-\left( \pi+1
\right)}\int_{0}^{2\pi}d\omega ~ \gamma(\omega)\frac{e^{i\omega
x}}{\left[z_{+}(\omega)\right]^{y}}.
\label{eq:finaldensity}
\end{equation}
This expression can be reduced to a compact form by a change of variable
$\omega\rightarrow 2q$ (for details see \ref{app:I2})  which yields
\begin{eqnarray}
\fl\phi\left( x,y \right)-\rho=\frac{2 \epsilon \rho}{
\left[2\pi-\epsilon\left( \pi+1 \right)\right]}\nonumber\\
\int_{0}^{\pi/2} \frac{\cos\left[ q\left( 2x+1
\right)\right] \cos q -\sin\left[ q\left( 2x+1\right)
\right]\sqrt{1+\sin^{2}q}}{\left(1+2\sin^{2}q+2\sin
q\sqrt{1+\sin^{2}q}\right)^{y}} dq.
\label{eq:finalsimp}
\end{eqnarray}
The difference in the density from the uniform profile
$\rho$ is proportional to the blockage strength $\epsilon$, and
vanishes when $\epsilon=0$. Also higher the bulk density $\rho$, more
pronounced is the effect.

\begin{figure}
\begin{center}
\includegraphics[width=10cm]{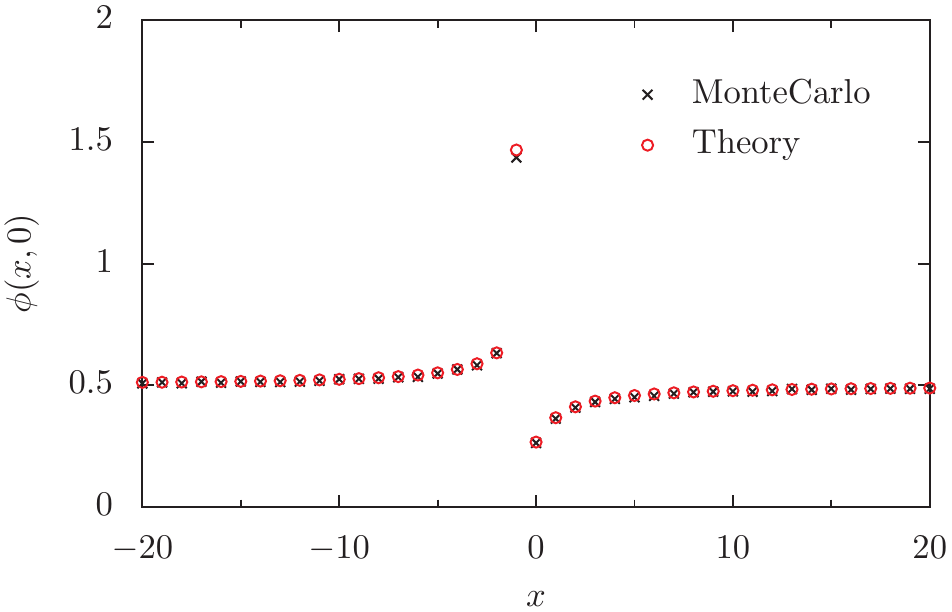}
\caption{A comparison of the density profile in
\eref{eq:finalsimp} with the Monte Carlo results at sites along the
driven lane. The $x$ is the site index on the
driven lane. For convenience we denote the site $(L-i,0)$ on the left of
the broken bond as $x=-i$. The broken bond is between $x=-1$th
and $x=0$th site, while the drive is along the positive $x$ direction.
\label{fig:fig4}
}
\end{center}
\end{figure}

To verify the relevance of the $L \rightarrow \infty$ limit to finite $L$ systems, the 
density profile is compared with numerical
data from Monte Carlo simulation in \Fref{fig:fig4}. The simulation is performed on a $100\times100$ lattice with a broken bond ($\epsilon=1$) between $(99,0)$
site and $(0,0)$ site. In the starting configuration $5000$ particles were
distributed randomly on the lattice leading to a global average density
$\rho=1/2$. We follow a random sequential update rule: in every time step, all the particles are updated exactly once following the
stochastic dynamics in \sref{sec:model}. In spite of the finite size of the lattice, the density
profile matches very well with the theoretical result in \eref{eq:finalsimp} which corresponds to the $L\rightarrow\infty$ limit. In \Fref{fig:fig4} we present only the results along
the driven lane.

\subsection{The asymptotic profile}
The density has a maximum at the site $(L-1,0)$ which is the left end of the slow
bond, whereas it is has a minimum at the site $(0,0)$ on the right of the slow
bond. Far from the slow bond the density approaches the global average
value $\rho$. The convergence to this value is slow, and most importantly the difference $\phi(x,y)-\rho$ decays algebraically as $1/\sqrt{x^2+y^2}$, in
all directions far from the slow bond, except along the diagonal $y=x$ where it decays as $1/(x^2+y^2)^{3/2}$. To show this power law tail, we analyze the solution \eref{eq:finalsimp} in three directions,
namely, along the driven lane ($y=0$), along the $x=0$ line, and also
along the line with slope $m$. For convenience we denote the integral in  \eref{eq:finalsimp} by
\begin{equation}
I(x,y)=\int_{0}^{\pi/2} \frac{\cos\left[ q\left( 2x+1
\right)\right] \cos q -\sin\left[ q\left( 2x+1\right)
\right]\sqrt{1+\sin^{2}q}}{\left(1+2\sin^{2}q+2\sin
q\sqrt{1+\sin^{2}q}\right)^{y}} dq.
\label{eq:I}
\end{equation}

\subsubsection*{Along the driven lane:}
For $y=0$ the integral $I(x,y)$ simplifies to
\begin{equation}
\fl \qquad I(x,0)=\int_{0}^{\pi/2}dq \cos\left[ \left( 2x+1 \right) q \right]\cos q
-\int_{0}^{\pi/2}dq \sin\left[ \left( 2x+1 \right) q
\right]\sqrt{1+\sin^{2} q}.
\end{equation}
It is easy to show that the first integral vanishes for all integer
values of $x$ except at $x=-1$ and $x=0$ where its value is
$\pi/4$. To evaluate the second integral, we make a change of variable with
$(2x+1)q=\xi$. Then the density profile yields,
\begin{eqnarray}
\phi(x,0)-\rho=&\frac{2\epsilon\rho}{2\pi-\epsilon(\pi+1)}\left[\frac{\pi}{4}\left(
\delta_{x,-1}+\delta_{x,0} \right)\right.\nonumber\\
&\left.-\frac{1}{2x+1}\int_{0}^{(2x+1)\pi/2}d\xi\sin\xi \sqrt{1+\sin^{2}\left(
\frac{ \xi}{2x+1}
\right)}\right].
\end{eqnarray}
The term inside the square root in the integrand is slowly varying, and for large
$\vert x\vert$ it remains almost constant while $\sin\xi$
completes a cycle. Then the integral can be approximated as
\begin{equation*}
\int_{0}^{\pi/2}d\xi\sin\xi+\sum_{n=1}^{x}\sqrt{1+\sin^{2}\left(
\frac{n\pi}{2x+1}
\right)}\int_{\frac{(2n-1)\pi}{2}}^{\frac{(2n+1)\pi}{2}}d\xi \sin\xi.
\end{equation*}
The first integral is equal to $1$ whereas the one inside the summation
is zero. This yields, for large $\vert x\vert$, the
power law tail,
\begin{equation}
\phi\left( x,0
\right)-\rho\simeq-\frac{2\epsilon\rho}{2\pi-\epsilon(\pi+1)}\times
\frac{1}{2x+1}.
\end{equation}

An interesting feature to note that, except for the two sites ($x=-1$ and $x=0$) the profile is anti-symmetric with respect to the slow bond (see \fref{fig:fig4}).

\subsubsection*{Along the $x=0$ line:}
Along this line, perpendicular to the driven lane, the integral in \eref{eq:I}
yields,
\begin{equation}
I(0,y)=\int_{0}^{\pi/2}dq\frac{\cos^{2}q-\sin
q\sqrt{1+\sin^{2}q}}{\left[1+2\sin^{2}q+2\sin
q\sqrt{1+\sin^{2}q}\right]^{y}}.
\end{equation}
The denominator in the integrand has its minimum value at $q=0$ and
it monotonically increases with increasing $q$ within the interval of
integration. Thus the contributions for large $y$ comes predominantly from
small $q$. Expanding around $q=0$ the integral can be approximated as
\begin{equation}
I(0,y)\simeq\int_{0}^{\pi/2}dq\left(\cos^{2}q-\sin
q\sqrt{1+\sin^{2}q}\right)e^{-2yq}.
\end{equation}
Due to the exponential damping, the leading contribution comes from
$q\lessapprox 1/2y$ where $e^{-2yq}\simeq1$ and the integral yields
\begin{equation}
I(0,y)\simeq\int_{0}^{1/2y}dq\left(\cos^{2}q-\sin
q\sqrt{1+\sin^{2}q}\right)\simeq
\frac{1}{2y}+\mathcal{O}(1/y^{2}).
\end{equation}
Thus the density far from the driven lane, approaches the global average value $\rho$ as
\begin{equation}
\phi\left( 0,y \right)-\rho\simeq\frac{2\rho\epsilon}{2\pi-(\pi+1)}\times
\frac{1}{2y}.
\end{equation}

\subsubsection*{Along the line with slope $m$:}
Consider a line $2y=m(2x+1)$ with $m$
being the slope. As a start, let both $m$ and $x$ are positive.
The integral in \eref{eq:I} yields
\begin{equation}
\fl \qquad I\left[ x,m\left(x+\frac{1}{2}\right)\right]=\int_{0}^{\pi/2}dq \frac{\cos\left[ q\left( 2x+1
\right)\right] \cos q- \sin
q\left( 2x+1 \right)\sqrt{1+\sin^{2}q}}{\left[1+2\sin^{2}q+2\sin
q\sqrt{1+\sin^{2}q}\right]^{m\left( x+1/2 \right)}}.
\label{eq:alongslop}
\end{equation}
Like the case discussed above, the integral can be approximated
by considering the contributions only from small $q$ whereby it
reduces to
\begin{eqnarray}
\fl I\left[ x,m\left(x+\frac{1}{2}\right)\right]\simeq\int_{0}^{\pi/2}dq\left\{\cos\left[ q(2x+1)\right]\cos q-\sin\left[ q(2x+1)
\right]\sqrt{1+\sin^{2}q}\right\}\nonumber\\
\qquad\qquad\qquad\qquad\qquad\qquad\qquad\qquad\qquad\qquad\qquad e^{-m(2x+1)q}.
\end{eqnarray}
Applying a change of variable $q(2x+1)=\xi$ and keeping only the
leading orders in $1/x$ the integral simplifies to
\begin{equation}
 I\left[ x,m\left(x+\frac{1}{2}\right)\right]\simeq\frac{1}{2x+1}\int_{0}^{(x+1/2)\pi}d\xi\left(\cos\xi-\sin\xi\right)e^{-m\xi}.
\end{equation}
The integral is now easy to compute, yielding the density difference
\begin{equation}	
	\phi\left[ x,m\left(x+\frac{1}{2}\right)
\right]-\rho\simeq\frac{2\rho\epsilon}{2\pi-(\pi+1)}\times
\frac{m-1}{(m^{2}+1)(2x+1)}.
\label{eq:33}
\end{equation}

Note that, the density difference changes sign as $m$ crosses
$1$. In fact, at $m=1$, \textit{i.e.}, along the line $2y=2x+1$, the
leading order term in the integral $I$ vanishes. By considering the higher
order contributions in $1/x$, it can be shown that the integral
for $m=1$ yields
\begin{equation}
I\left[ x,\left(x+\frac{1}{2}\right)\right]\simeq -\frac{1}{2\left( 2x+1 \right)^{3}}.
\end{equation}
For more details see \ref{app:excl}, where the integral $I$ can be extracted from \eref{eq:II3} by substituting $\rho=0$. Thus, along the diagonal, the density profile
\begin{equation}
\phi\left( x,x+\frac{1}{2}
\right)-\rho\simeq-\frac{2\rho\epsilon}{2\pi-(\pi+1)}\times
\frac{1}{2(2x+1)^{3}}.
\end{equation}

The analysis can be easily extended for negative $x$ and $m$. The expression for the profile \eref{eq:33} remains unchanged.

\subsubsection*{In polar coordinate:}
It is instructive to express the density in terms of the polar coordinates $\theta=\arctan(y/x)$ and $r=\sqrt{x^2+y^2}$. At large distances, the density profile,
\begin{equation}
\phi(r,\theta)-\rho\simeq\frac{\sqrt{2}\rho\epsilon}{2\pi-(\pi+1)}\frac{\sin(\theta-\frac{\pi}{4})}{r}.
\label{eq:densitynonpolar}
\end{equation}
For $\theta=\pi/4$ where the expression vanishes, the
profile is determined by the sub-leading contribution decaying as
$1/r^{3}$.

The slow decay of the density profile is also reflected in the induced
current. As the particles are independent of each other, the current is
due to diffusion except along the driven lane. The particle
current at any site $(x,y)$ can be expressed in terms of the local
density profile as
\begin{equation}
\mathbf{J}\left( x,y
\right)=\left[ 1-\epsilon\delta_{x,-1} \right]\phi(x,y)\mathbf{\hat
x}-\frac{\partial \phi}{\partial y}\mathbf{\hat y}\textrm{~ ~ ~  ~ ~ ~ for
\ensuremath{y=0},}
\end{equation}
\begin{equation}
\mathbf{J}\left( x,y \right)=-\mathbf{\nabla} \phi\left(
x,y\right)\textrm{  ~ ~ ~ ~ ~ ~ ~ ~  ~ ~ ~ ~ ~ ~ ~ ~elsewhere,}
\end{equation}
where $\mathbf{\hat x}$ and  $\mathbf{\hat y}$ are the unit vectors along
the $x$ and $y$ directions, respectively. From the density profile in \eref{eq:densitynonpolar}, it is clear that far from the
slow bond, the particle current decreases algebraically. Particularly,
in terms of the polar coordinates the induced current in the bulk,
\begin{equation}
\mathbf{J}(r,\theta)\simeq\frac{\sqrt{2}\rho\epsilon}{2\pi-(\pi+1)}\times\frac{\sin(\theta-\frac{\pi}{4})\mathbf{\hat{r}}-\cos(\frac{\theta-\pi}{4})\boldsymbol{\hat{\theta}}}{r^{2}},
\end{equation}
where $\mathbf{\hat{r}}$ and $\boldsymbol{\hat{\theta}}$ are the unit
vectors along the radial and the angular directions.

\section{Exclusion interaction\label{sec:excl}}
Most of the qualitative features of the stationary profile remain
unchanged from the independent particle case. In the absence
of the slow bond the surface drive does not affect the density profile
which remains uniform through out the lattice. However, introducing a slow bond on the driven lane makes the particles queue behind the
bond, resulting in a density gradient that
propagates far inside the bulk and induces diffusive current.
The difference in density from the uniform profile decays algebraically
with the distance $r$ from the slow bond. The decay
exponent is same as that in the independent particles case.

We now proceed to derive the density profile from the rate equations at the stationary state. Let $ n_{t}(x,y)$ be the occupation variable of site $(x,y)$ at time $t$ which takes value $1$ if the site is occupied and $0$ if there is no particle. The density at time $t$ is obtained by performing ensemble
average, $\phi_{t}(x,y)=\langle n_{t}(x,y)\rangle$. Following the
dynamical rules in \sref{sec:model}, it is easy to show that the time evolution of $\phi_t(x,y)$ is
\begin{equation}
\fl\frac{\partial \phi_{t}\left( x, y\right)}{\partial t} = \left\{
  \begin{array}{l l}
  \Delta\phi_{t}(x,y) &\textrm{          for \ensuremath{y>0},}\\
  ~ & ~\\
  \phi_{t}(x-1,0) +\phi_{t}(x,1)-2\phi_{t}(x,0)&\\
  -\langle n_{t}(x,0) n_{t}(x-1,0)\rangle+\langle n_{t}(x,0) n_{t}(x+1,0)\rangle\nonumber&\\
  -\epsilon\langle n_{t}(L-1,0)(1- n_{t}(0,0))\rangle[\delta_{x,0}-\delta_{x,L-1}]&\textrm{for \ensuremath{y=0},}\\
  \end{array} \right.\nonumber
\end{equation}
where $\Delta$ is the discrete Laplacian defined in
\eref{eq:dl}. In the stationary state, the average profile does not depend on
time, and the time derivative vanishes. Then the
equation governing the profile yields
\begin{eqnarray}
\qquad\qquad \Delta\phi(x,y)=&0 \qquad \qquad \qquad\textrm{ for  \ensuremath{y>0},}\nonumber\\
\fl\quad\phi(x-1,0)+\phi(x,1)-2\phi(x,0)=&\langle n(x,0)n(x-1,0)\rangle-\langle n(x,0) n(x+1,0)\rangle\nonumber\\
&+\epsilon\langle
n(L-1,0)(1-n(0,0))\rangle\left[\delta_{x,0}-\delta_{x,L-1}\right]\nonumber\\
&\qquad\qquad\qquad\quad\textrm{for \ensuremath{y=0}},
\label{eq:masterexcl}
\end{eqnarray}
where we dropped the time index in both $\phi$ and $ n$.

In the absence of the slow bond, it is easy to show that in the steady state all configurations are
equally probable. This directly implies that the density profile is
uniform. It is worth mentioning that, a very different behavior was observed in a related work
\cite{DICKMAN,POTIGUAR} with \textit{nearest neighbor} exclusion interaction,
where a shear drive rearranges the average population of the lanes.

For non-zero $\epsilon$, the slow bond breaks the translation invariance,
and the uniform density profile is no longer a stationary state.
It is important to note that, like in the case of independent particles, the average population in each lane remains same, $L^{-1}\sum_{x}\phi(x,y)=\rho$.
This feature will be important in our derivation of the stationary profile.

\subsection{Mean-field analysis}
An exact analysis of the stationary profile is hard as the
\Eref{eq:masterexcl} involves two point correlations of the occupation
variables which in turn depends on higher order correlations. It is almost impossible to circumvent this hierarchy. However, the rate equations can be simplified within a mean field
approximation, \textit{i.e}, by imposing factorization assumption
\begin{equation}
\langle  n(x,0)  n(x+1,0) \rangle = \phi(x,0)\phi(x+1,0).
\end{equation}
Such mean-field approximation have been successfully
applied to a variety of driven diffusive systems, see
\textit{e.g.} \cite{MUKAMEL,BLYTHE}.
Within mean field approximation, the stationary \Eref{eq:masterexcl} becomes
\begin{eqnarray}
\qquad\qquad\Delta\phi(x,y)&=& 0\quad \quad \quad \quad \quad \quad \quad \quad ~\textrm{ for \ensuremath{y>0},}\label{eq:masterexcl1}\\
\fl\phi(x-1,0)+\phi(x,1)-2\phi(x,0)&=&\phi(x,0)\phi(x-1,0)-\phi(x,0)\phi(x+1,0)\nonumber\\
&&+\epsilon\phi(L-1,0)(1-\phi(0,0))[\delta_{x,0}-\delta_{x,L-1}]\nonumber\\
&&\quad \quad \quad \quad \quad \quad \quad \quad \quad \textrm{ for }y=0.
\label{eq:masterexcl2}
\end{eqnarray}
The boundary condition is that at large distances from the slow bond the
density converges to a global average value $\rho$.

The solution is still difficult because of the non-linearity.
We proceed by using a perturbative expansion of the density $\phi(x,y)$ in powers of $\epsilon$. We assume that $\phi(x,y)$ is Taylor expandable around $\epsilon=0$ state where
the density profile is uniform, $\rho$ at all
sites. Let
\begin{equation}
\phi(x,y)=\rho+\sum_{p=1,2,3}^{\infty}\epsilon^{p}\alpha_{p}(x,y).
\label{eq:perturbative}
\end{equation}
The advantage is that to each order in the expansion the corresponding equations become linear. In this section, we analyze only the linear
order term which captures the power-law tail of the density profile. The higher order terms do not change the power-law. A detailed analysis of the higher order terms is presented in the \ref{app:higher order terms}.

At large distances away from the slow bond, the density reaches a uniform
profile equal to $\rho$. This implies that $\alpha_{1}(x,y)$ must vanish
at large distances from the slow bond. Substituting the perturbative expansion
into \Eref{eq:masterexcl1}-\eref{eq:masterexcl2} yields:
\begin{equation}
\nabla^{2}\alpha_{1}(x,y)=0 \qquad \qquad \qquad \qquad \qquad\textrm{for \ensuremath{y>0}},
\label{eq:alpha one master non-zero y}
\end{equation}
and,
\begin{eqnarray}
\alpha _1(x-1,0)+\alpha _1(x,1)-2\alpha _1(x,0)+\rho \left(\alpha
_1(x+1,0)-\alpha _1(x-1,0)\right) \nonumber\\
\qquad\qquad=\rho(1-\rho)\left(\delta _{x,0}-\delta
_{x,-1}\right) \qquad\textrm{for \ensuremath{y=0}}.
\label{eq:alpha one master zero y}
\end{eqnarray}
This is a set of coupled linear equations and can be solved in a
way similar to the solution of \Eref{eq:masternon1}-\eref{eq:masternon2}, in the
independent particle case. Let, the Fourier modes are defined as
\begin{eqnarray}
g_{1}(n,y)=&\frac{1}{L}\sum_{x=0}^{L-1}\alpha_{1}\left( x,y
\right)e^{-i \omega_n x}
\label{eq:fourier app}
\end{eqnarray}
where $\omega_n=2\pi n/L$ with $n=0,1,\cdots,L-1$. The subscript
$1$ in the Fourier amplitudes indicates that the calculations are for
the order one term in $\epsilon$. In terms of these
Fourier amplitudes, the equations \eref{eq:alpha one master non-zero y} and
\eref{eq:alpha one master zero y} yield
\begin{equation}
g_{1}(n,y+1)=-g_{1}(n,y-1)+2\left[ 2-\cos(\omega_{n}) \right]g_{1}(n,y)
\textrm{~ ~  for \ensuremath{y>0}},
\label{eq:fmode app1}
\end{equation}
and
\begin{equation}
g_{1}(n,1)=\left[ 2(1-i\rho\sin\omega_{n}) -
e^{-i\omega_{n}}\right]g_{1}(n,0)+\frac{\rho(1-\rho)}{L}\left(
1-e^{i\omega_{n}}\right),
\label{eq:fmode app2}
\end{equation}
respectively.

By definition in \eref{eq:fourier app}, $g_{1}(0,y)=L^{-1}\sum_{x}\alpha_{1}(x,y)$. Given that the average $L^{-1}\sum_{x}\phi(x,y)=\rho$ for all $y$, the amplitude of the zeroth Fourier mode must vanish, $g_{1}(0,y)=0$, for all $y$. Taking this into account, the inverse Fourier
transformation yields the formal solution
\begin{equation}
\alpha_{1}(x,y)=\sum_{n=1}^{L-1}g_{1}(n,y)e^{i \omega_{n}x}.
\label{eq:invfou app}
\end{equation}

The amplitudes $g_{1}(n,y)$ of the Fourier modes can be calculated
iteratively using the recurrence relation \eref{eq:fmode app1}. A systematic approach of solving this
recurrence relation is using the generating function, defined as
\begin{equation}
G_{1}(n,z)=\sum_{y=1}^{\infty}g_{1}(n,y)z^{y}.
\label{eq:Genfunc app}
\end{equation}
A similar method is used for solving the profile for the independent
particles. Using \eref{eq:fmode app2} the generating function yields a similar expression as in \eref{gz},
\begin{equation}
G_{1}(n,z)=\frac{g_{1}(n,1)-zg_{1}(n,0)}{(z-z_{-})\left(
z-z_{+}
\right)}\times z,
\label{eq:gen fncexprpp}
\end{equation}
for $n>0$ with same $z_{\pm}$ as in \eref{eq:zpm}.

A pole cancelling argument used earlier for the independent particles case, simplifies the expression for
the generating function and can be used to determine the Fourier modes $g_{1}(n,y)$. From \eref{eq:fmode app1},
it is clear that $\vert g_{1}(n,y)\vert \le L^{-1} \sum_{x}\vert\alpha_{1}(x,y)\vert$, and as at large distances from
the driven lane the density is close to the uniform profile, $\vert
g_{1}(n,y)\vert$ is finite for $y\rightarrow \infty$.
Then, the series in \eref{eq:gen fncexprpp} has a radius of convergence $\vert z \vert <1$. Then, the generating
function should not have any pole within the unit circle around the origin on the complex $z$-plane. This yields
\begin{equation}
g_{1}(n,0)=\Gamma_{\rho}(\omega_{n})\left[\frac{\rho(1-\rho)}{L}\left(
e^{i\omega_n}-1\right)\right],
\label{eq:gn0 app}
\end{equation}
where
\begin{equation}
\Gamma_{\rho}\left( \omega_{n}
\right)=\frac{1}{\left[\sqrt{\left(
2-\cos\omega_{n} \right)^{2}-1}\right]+i\left( 1-2\rho
\right)\sin\omega_{n}}.
\label{eq:Gamma excl}
\end{equation}
In addition, the Fourier modes for any $y$, yields
\begin{equation}
g_{1}(n,y)=\frac{g_{1}\left( n,0 \right)}{z_{+}^{y}}.
\label{eq:gny app}
\end{equation}
Incorporating these expressions of $g_{1}(n,y)$ in \eref{eq:gn0 app}
and \eref{eq:gny app} to the solution \eref{eq:invfou app},
$\alpha_{1}(x,y)$ can be written as
\begin{equation}
\alpha_{1}(x,y)=\frac{\rho(1-\rho)}{L}\sum_{n=1}^{L-1}\left(
e^{i\omega_{n}}-1
\right)\Gamma_{\rho}(\omega_{n})\frac{e^{i\omega_{n}x}}{\left[z_{+}(\omega_{n})\right]^{y}}.
\label{eq:genexpr app}
\end{equation}
Then, finally the solution for the density profile can be expressed as,
\begin{equation}
\phi(x,y)-\rho=\epsilon \rho(1-\rho)\frac{1}{L}\sum_{n=1}^{L-1}\frac{\gamma_{\rho}(\omega_{n})\times
e^{i\omega_{n}x}}{\left[z_{+}(\omega_{n})\right]^{y}}+\mathcal{O}\left(\epsilon^{2} \right),
\label{eq:genexpr app2}
\end{equation}
where
\begin{equation}
\gamma_{\rho}(\omega_{n})=(e^{i\omega_{n}}-1)\Gamma_{\rho}(\omega_{n}).
\label{eq:small gamma rho}
\end{equation}
Notice that for $\rho=0$, the $\gamma_{\rho}(\omega)$ is same as
$\gamma(\omega)$ defined in \eref{eq:gamma} for the independent particle case.

In the $L\rightarrow \infty$ limit, the
profile can be expressed in terms of continuous variables $\omega \equiv \omega_n$, and the
summation over $n$ can be approximated by an integration, yielding,
\begin{equation}
\phi\left( x,y
\right)=\rho+\epsilon \rho (1-\rho)\int_{0}^{2\pi} \frac{d\omega}{2\pi} \frac{\gamma_\rho(\omega)e^{i\omega x}}{\left[
z_{+}\left( \omega \right)
\right]^{y}}+\mathcal{O}\left( \epsilon^{2} \right),
\label{eq:density suma}
\end{equation}

It will be shown in the following section that this leading order term in $\epsilon$ already captures the algebraic profile at large distances. It turns
out that, the higher order contributions do not alter the power-law tail of
the profile; it only changes the overall amplitude. By calculating these higher order terms in the \ref{app:higher order terms} we shall show that at large distances the density profile,
\begin{equation}
\phi\left( x,y
\right)= \rho+ \epsilon \phi(-1,0)\left[ 1-\phi(0,0)\right]\int_{0}^{2\pi}
\frac{d\omega}{2\pi} \frac{\gamma_\rho(\omega)e^{i\omega x}}{\left[
z_{+}\left( \omega \right)
\right]^{y}}.
\label{eq:genexpr app3}
\end{equation}
When compared with the profile \eref{eq:finaldensity} in the independent particles case, the integrand above differs in the
factor $\gamma_{\rho}(\omega)$ which is a function of $\rho$.

\subsection{The asymptotic profile}
The density profile decays algebraically to the uniform
value $\rho$, at large distances. In most directions away from the slow
bond, the difference decays as $1/r$ with $r$ being the distance from the slow bond. Only at a certain angle the decay is faster as $1/r^3$.

To analyze the profile, we write the expression in \eref{eq:genexpr app3} as
\begin{eqnarray}
\fl \qquad \phi(x,y)-\rho= \left(\frac{ \epsilon \phi(-1,0)\left[ 1-\phi(0,0)\right]}{\pi}\right)\int_{0}^{\pi/2}dq\frac{1}{\left[ 1-2\rho(1-\rho)\cos^{2}q
\right]}\nonumber\\
\qquad\qquad \times\frac{(1-2\rho)\cos\left[
(2x+1)q \right]\cos q-\sin\left[(2x+1)q
\right]\sqrt{1+\sin^{2}q}}{\left[ 1+2\sin^{2}q+2\sin q\sqrt{1+\sin^{2}q}
\right]^{y}}.
\label{eq:90}
\end{eqnarray}
For details see \ref{app:I2}.
Notice that the profile is composed of a symmetric (with cosine term) part and an anti-symmetric part (with sine term) under $x\rightarrow -(x+1)$. It can be shown that the symmetric part falls of exponentially with increasing $x$, and at large distances the profile effectively becomes anti-symmetric (see \Fref{fig:ExNear}).

The analysis of the asymptotic profile in different directions is similar to that in the independent particles case. A detailed
calculation is deferred to the \ref{app:excl}.
\subsubsection*{Along the driven lane:}
\begin{equation}
\phi(x,0)-\rho \simeq - \left(\frac{ \epsilon \phi(-1,0)\left[ 1-\phi(0,0)\right]}{\pi\left[1-2\rho(1-\rho)\right]}\right)\frac{1}{2x+1}
\label{eq:phix}
\end{equation}
\subsubsection*{Along $x=0$ line:}
For $\rho\ne1/2$
\begin{equation}
\qquad\phi(0,y)-\rho\simeq\left(\frac{ \epsilon \phi(-1,0)\left[ 1-\phi(0,0)\right]}{\pi\left[1-2\rho(1-\rho)\right]}\right)\times\frac{
1-2\rho }{2y},
\label{eq:phiy}
\end{equation}
and for $\rho=1/2$
\begin{equation}
\qquad\phi(0,y)-\rho\simeq-\left(\frac{ \epsilon \phi(-1,0)\left[ 1-\phi(0,0)\right]}{\pi\left[1-2\rho(1-\rho)\right]}\right)\times\frac{
1 }{8y^2},
\end{equation}

\subsubsection*{Along a line of slope $m$:}
For $m(1-2\rho)\ne1$,
\begin{equation}
\fl \qquad\phi\left[x,m(x+1/2)\right]-\rho=
\frac{ \epsilon \phi(-1,0)\left[ 1-\phi(0,0)\right]}{\pi\left[1-2\rho(1-\rho)\right]}\times\frac{m(1-2\rho)-1}{m^{2}+1}\times\frac{1}{2x+1},
\end{equation}
and for $m(1-2\rho)=1$,
\begin{equation}
\fl \qquad\phi\left[x,m(x+1/2)\right]-\rho=-
\frac{ \epsilon \phi(-1,0)\left[ 1-\phi(0,0)\right]}{2\pi\left[1-2\rho(1-\rho)\right]}\times\frac{\left[1+2\rho(1-\rho)\right](1-2\rho)^{4}}{\left(2x+1\right)^3}.
\end{equation}

\subsubsection*{In polar coordinate:}
Like in the independent particles case, the above asymptotic results can be put together in a simple expression in the polar
coordinates $r=\sqrt{x^2+y^2}$ and $\theta=\arctan(y/x)$. For
large $r$, and $0\le \theta\le \pi$,
\begin{equation}
\phi(r,\theta)-\rho\simeq\frac{ \epsilon \phi(-1,0)\left[ 1-\phi(0,0)\right]}{\pi\sqrt{2-4\rho(1-\rho)}}\times
\frac{\sin\left(\theta-\Theta \right)}{r},
\label{eq:phipolar}
\end{equation}
where $\Theta=\pi/2-\arctan[(1-2\rho)/\sqrt{1+(1-2\rho)^2}]$. For
$\theta=\Theta$ where the leading term vanishes, the profile decays as
$1/r^3$. The angle depends on the average density
$\rho$, and varies from $\pi/4$ to $3\pi/4$ as $\rho$ is increased
from $0$ to $1$. We recall that, for the independent particles case, this angle is at $\theta=\pi/4$ [see \Eref{eq:densitynonpolar}].

Notice, that as $\rho\rightarrow (1-\rho)$ the angle
$\Theta\rightarrow \pi-\Theta$. This results in a symmetry of the
profile
\begin{equation}
\eqalign\phi_{1-\rho}(r,\theta)=1-\phi_{\rho}(r,\pi-\theta).
\end{equation}
This is a direct consequence of the particle-hole (empty site) equivalence in the exclusion
process: A particle jumping across a bond to an empty site can
also be considered as a hole moving in the opposite direction.
\begin{figure}
\begin{center}
\includegraphics[width=12cm]{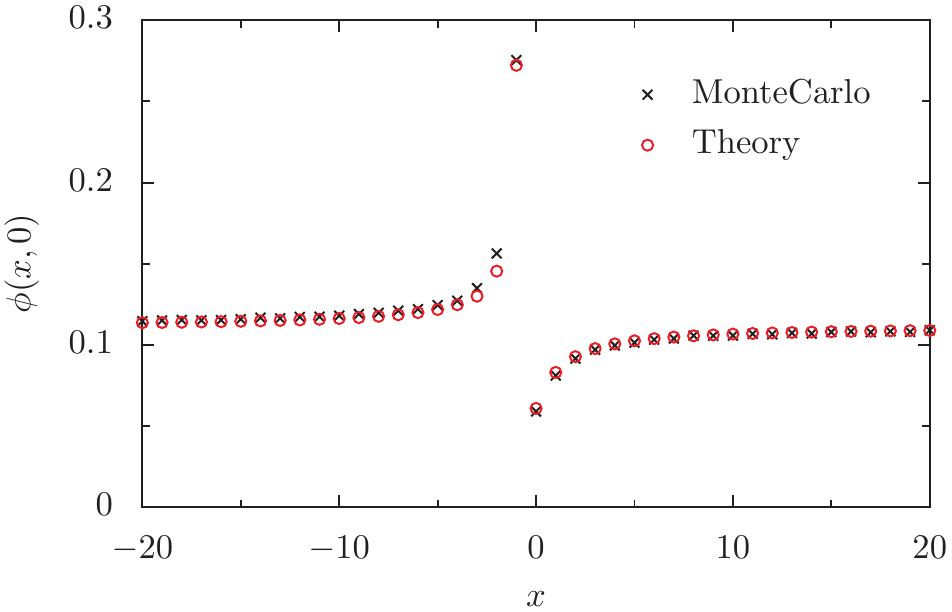}
\caption{A comparison of the computer simulation results of the stationary
density profile $\phi(x,0)$ in the exclusion case with the solution \eref{eq:90}  obtained using mean-field approximation. The slow bond is between $x=-1$ and $x=0$ th sites.}
\label{fig:ExNear}
\end{center}
\end{figure}

The long-range density profile, induces diffusive current in the bulk. Within the mean-field approximation the stationary state current can be expressed as
\begin{equation}
\fl \qquad\qquad\mathbf{J}(x,0)=\phi(x,0)\left[ 1-\phi(x+1,0) \right]\left[ 1-\epsilon
\delta_{x,-1}\right]\mathbf{\hat x}-\frac{\partial \phi(x,y)}{\partial
y}\mathbf{\hat y}\qquad\textrm{ for \ensuremath{y=0},}
\end{equation}
\begin{equation}
\fl \qquad\qquad\mathbf{J}(x,y)=-\boldsymbol{\nabla}\phi\left( x,y
\right)\qquad\qquad\qquad\qquad\qquad\qquad\qquad\qquad\quad\textrm{ elsewhere.}
\end{equation}
The $\mathbf{\hat x}$ and $\mathbf{ \hat y}$ are the unit vectors in
the $x$ and $y$ directions, respectively. It is clear from the
algebraic decay of the density profile that the induced current in the
bulk decays as $1/r^2$ in all directions, with the distance away from
the slow bond. Particularly, in the polar coordinates, using the density profile
in \eref{eq:phipolar}, the induced current in the bulk can
be expressed as
\begin{equation}
\mathbf{J}(r,\theta)\simeq\left(\frac{ \epsilon \phi(-1,0)\left[ 1-\phi(0,0)\right]}{\pi\sqrt{2-4\rho(1-\rho)}}\right)\times\frac{\sin(\theta-\Theta)\mathbf{\hat{r}}-\cos(\theta-\Theta)\boldsymbol{\hat{\theta}}}{r^{2}},
\end{equation}
where $\mathbf{\hat r}$ and $\mathbf{\hat \theta}$ are the unit
vectors along the polar coordinates $r$ and $\theta$, respectively.

\begin{figure}
\begin{center}
\includegraphics[width=12cm]{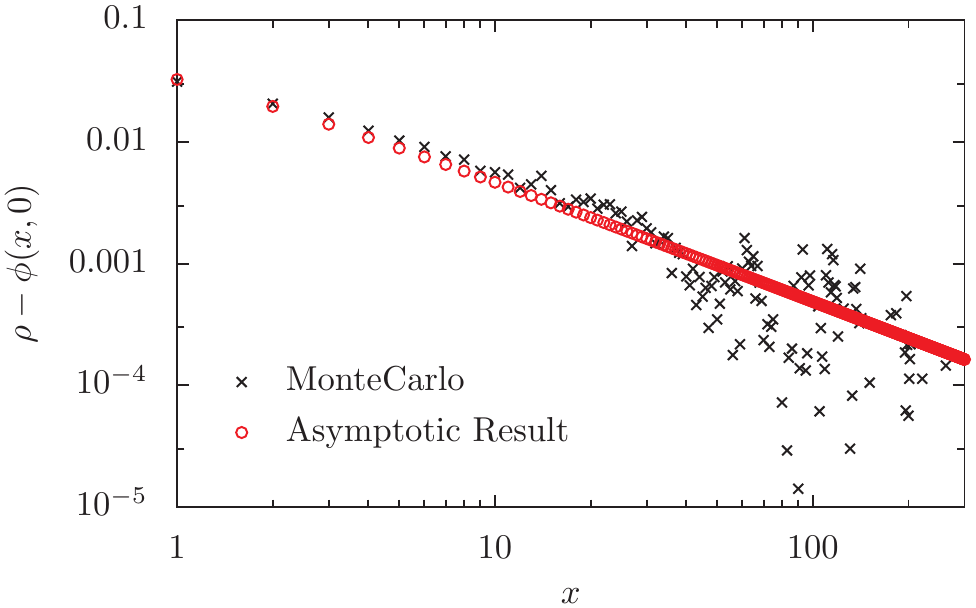}
\caption{The density difference at sites on the right of the broken bond along the driven lane. The density approaches the global average density
$\rho=1/9$ in the bulk, with the difference decaying inversely with
the distance $x$ away from the broken bond.}
\label{fig:ExMCx}
\end{center}
\end{figure}

\subsection{Numerical results}
Although the profile \eref{eq:90} is derived within a mean-field approximation, it describes the numerical result quite well.
We demonstrate this by comparing with the profile generated by
a Monte Carlo simulation on a $600\times 1200$ lattice with an average density $\rho=1/9$.
We consider the case with $\epsilon=1$, \textit{i.e.}, no jumps are allowed across the bond between $(L-1,0)$
and $(0,0)$ sites. In the simulation the particles are evolved
following a random sequential update,
where in every step of the iteration a site is chosen at random.
If the site is occupied, the particle is transferred to a randomly chosen
nearest neighbor site only when the latter is empty. One Monte Carlo time step consists of $LM$ such moves. Starting with a random distribution of particles, the system is evolved for a long time to
ensure that a stationary state distribution is reached. The density is averaged over one million configurations at intervals of $100$ Monte Carlo time steps.

The profile along the driven lane around the broken
bond is shown in \fref{fig:ExNear}. For brevity we denote the sites
$x=L-i$ on the left of the broken bond as $x=-i$. The theoretical results of the density
are calculated from the mean-field solution by substituting $y=0$ and $\epsilon=1$ in the expression in \eref{eq:90}. The integration is evaluated numerically. Although the expression \eref{eq:90} is applicable only at large distances, it describes the Monte Carlo result quite well, even near the broken bond.

\begin{figure}
\begin{center}
\includegraphics[width=12cm]{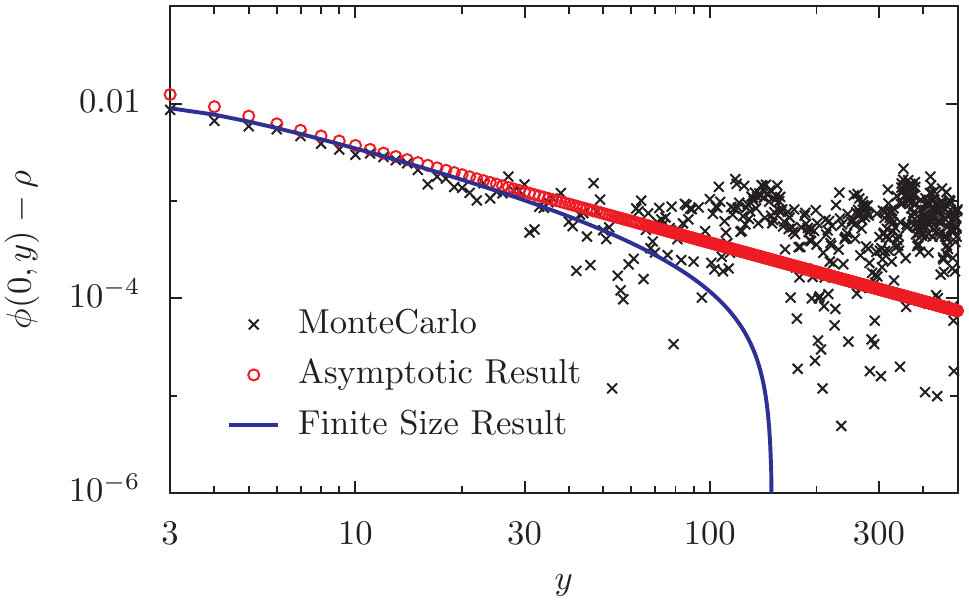}
\caption{The power law tail of the difference in density from its
global average value $\rho$, with the distance away from the driven
lane. The Asymptotic result is obtained from the mean-field solution \eref{eq:phiy},
whereas the finite size result is generated by numerical iteration
of the mean-field equation \eref{eq:masterexcl1}-\eref{eq:masterexcl2} on a finite system.}
\label{fig:ExMCy}
\end{center}
\end{figure}

The power-law decay of the profile can be seen in \fref{fig:ExMCx} where the profile along the boundary lane on the right of the slow
bond is plotted on a log-log scale. The $x$ range extends from the neighbor of the slow bond to the half of the system length. The result is consistent with
the $1/x$ decay in the asymptotic solution in \Eref{eq:phix}. The large scattering of the data at the tail of the profile is due to averaging over finite number of configurations ($10^{6}$).

The second direction where we compare the results is perpendicular
to the driven lane along the $x=0$ line. The Monte Carlo simulation
data and the results derived from the asymptotic expression in
\eref{eq:phiy} are plotted in
\fref{fig:ExMCy}. The saturation at the tail of the Monte Carlo data
is due to the finite sample size on which the data is averaged: the number fluctuation is comparable to the mean value. It is expected that when averaged over larger sample size the density difference would vanish at large distances. In fact, theoretically, on a finite
system size the power-law decay of the density difference is valid only up to a distance $\sim L$, and beyond this it decays exponentially. This is shown in \Fref{fig:ExMCy} by a continuous line which is obtained by numerical solution of the mean-field equation \eref{eq:masterexcl1}-\eref{eq:masterexcl2} on a $600\times 1200$ lattice. This is easy to understand from the fact that the density difference $\phi(x,y)-\rho$ satisfies Laplace's equation in the bulk,
whose solution for finite $L$ can be written as
\begin{equation}
\phi(x,y)-\rho=\sum_{n=0}^{L-1}\left[ A_{n} \cos\frac{2n\pi x}{L} + B_{n}
\sin\frac{2n\pi x}{L} \right]\exp\left( -\frac{2n\pi y}{L} \right),
\end{equation}
where $A_n$ and $B_n$ are the amplitudes of the Fourier modes.
It is clear that beyond $y\sim L$ the right hand
side decays exponentially.

\section{Electrostatic correspondence \label{sec:electro}}
The algebraic decay of the density profile is simple to understand from an electrostatic analogy. A similar correspondence proved useful in a related study of a lattice gas with localized bulk drive \cite{SMM,SMM3}.

Let us first consider the case of independent particles. For simplicity, we
consider the slow bond between the sites $(-1,0)$ and $(0,0)$, on a semi-infinite lattice with $x$ ranging from
$-\infty$ to $\infty$ and $y$ from $0$ to $\infty$. In the continuum limit, the stationary state equation in
\eref{eq:masternon1}-\eref{eq:masternon2} yields
\begin{eqnarray}
	\nabla^{2}\phi(x,y)&=&0\qquad\qquad\textrm{for all}\quad y>0,\\
	-\partial_y \phi(x,y)&\simeq &\epsilon \phi(-1,0)\left[\delta(x+1)-\delta(x)\right]-\partial_x \phi(x,y) \qquad~~\textrm{for}\quad y=0,
\end{eqnarray}
the $\delta(x)$ being the Dirac delta function.
In analogy with electrostatic, $\phi(x,y)$ is the potential on the upper half-plane due to a line charge density $\sigma(x)=-\partial_x \phi(x,0)$ at the boundary $y=0$ and a dipole of moment $\epsilon \phi(-1,0)$ at the origin.

As the charge density itself a function of the potential, the solution has to be
determined self-consistently. It is not difficult to see that a potential which has a dipole profile at large distances, is a consistent solution. It is consistent with the dipole at the origin. On the other hand,
as the potential $\phi(x,y)$ jumps discontinuously across the slow bond (see \Fref{fig:density}), the
$\sigma(x)$ consists of a positive charge at the origin and a distributed negative charge elsewhere. It is easy to
verify that the total amount of the charge $\sigma(x)$ summed over $x$ is zero.
Moreover, corresponding to the dipole solution, the charge $\sigma(x)$ decays as $1/x^2$ and it generates a quadrupolar potential which is sub-dominant to the contribution from the dipole at the origin. Then the large distance profile is that of a dipole potential, as we found in our exact derivation in the \eref{eq:finalsimp}.

In the case with exclusion interactions, from \eref{eq:masterexcl2}, the dipole at the origin is of moment $\epsilon\phi(-1,0)[1-\phi(0,0)]$ whereas the charge
\begin{equation*}
	\sigma(x)=-\left[ 1+2\phi(x,0) \right]\partial_x \phi(x,y).
\end{equation*}
Again, the total charge of $\sigma(x)$ is zero, and it generates a quadrupolar
potential at large distances. Then the potential due to the dipole at the origin
determines the large distance profile, as found in the solution of the
mean-field equation in \sref{sec:excl}. 

\section{Summary \label{sec:summary}}
One of the motivations for our work is to study how a shear flow at the boundary can induce current deep inside the bulk. In a fluid medium, this
non-local effect is not surprising as the particles due to their momentum carry
the directional information as they move away from the sheared layer. One simple example is the Couette
flow in viscous liquids, where a steady flow at the surface
induces current in the bulk whose amplitude decays linearly with the distance
from the sheared layer \cite{MUNSON}. Another example is provided by experimental \cite{DERKS} and theoretical \cite{SMITH1,SMITH2}
studies of the effect of shear drive on a fluctuating interface placed away from the boundary.
In the present paper, we showed that even in absence of the momentum degree of freedom (fluid with low Reynolds number), a similar non-local current can be induced, when there is a blockage at the sheared layer. Essentially, due to the particle conservation, the blockage at the boundary generates a non-local
density gradient across the system which in turn induces non-vanishing diffusive
current. We demonstrated this in a simple lattice-gas model with diffusive transport in
the bulk and shear flow at the boundary, along with a slow bond. We showed that in presence of the slow bond the density profile decays as
$1/y$ with the distance away from the shear layer. As a result, the diffusive
current decays as $1/y^{2}$.

We expect that the $1/y^2$ decay of the current is quite general and holds for
arbitrary local inter-particle interactions. Typically in a diffusive system, away from criticality, the
large scale properties of the conserved density field is effectively
described by a diffusion equation
\begin{equation}
\frac{\partial \phi_t(x,y)}{\partial t}\simeq -D
\nabla^2\phi_t(x,y),
\end{equation}
with $D$ being a diffusion constant. Then in the stationary state the
$\phi(x,y)$ follows the Laplace's equation. On the semi-infinite upper half-plane, as in our problem, a shear drive at the boundary induces a profile $\phi(x,0)$ at $y=0$. The solution of the Laplace's equation with this boundary condition can in general be written as
\begin{equation}
\phi(x,y)-\rho=\frac{y}{\pi}\int_{-\infty}^{\infty}dx'\frac{\phi(x',0)-\rho}{y^2+(x'-x)^2},
\end{equation}
where $\phi(x,y)$ approaches $\rho$ far from the boundary. This is known as
the Poisson formula. It is evident that
for a non-uniform profile $\phi(x,0)$ at the boundary, the difference $\phi(x,y)-\rho$ decays as
$1/y$. As a result the diffusive current decays as $1/y^{2}$. These results can be
generalized in higher dimensions, $d$, as well, where the induced current is found to decay as
$1/r^{d}$ with the perpendicular distance $r$ away from the shear drive.

\ack
We thank A. Bar, O. Cohen, M.R. Evans, O. Hirschberg, and S. Prolhac for helpful discussions. The support of the Israel Science Foundation
(ISF) and the Minerva Foundation with funding from the Federal Ministry of
Education and Research is gratefully acknowledged. S.N. Majumdar acknowledges support by ANR grant 2011-BS04-013-01 WALK-MAT and in part by the Indo-French Centre for the Promotion of Advanced Research under Project 4604-3.

\appendix
\section{Integration in \eref{eq:selfconsistencycont} \label{app:I1}}
To perform the integration
\begin{equation}
\mathcal{I}_{1}=\frac{1}{2\pi}\int_{0}^{2\pi}d\omega\gamma(\omega)\e^{-i\omega
}=\frac{1}{2\pi}\int_{0}^{2\pi}d\omega\frac{1-\e^{-i\omega
}}{\sqrt{(2-\cos\omega)^{2}-1}+i\sin\omega},
\end{equation}
consider a change of variable $\omega\rightarrow 2q$. In terms of
$q$, the integral reduces to,
\begin{equation}
\mathcal{I}_{1}=\frac{1}{\pi}\int_{0}^{\pi}dq \frac{e^{-iq}}{\cos
q-i\sqrt{1-\sin^{2}q}},
\end{equation}
which in terms of trigonometric functions yields
\begin{eqnarray}
\mathcal{I}_{1}=&\frac{1}{2\pi}\int_{0}^{\pi}dq\left[ \cos^{2} q +\sin
q\sqrt{1+\sin^{2}q}\right]\nonumber\\
&\quad\quad\quad\quad-\frac{i}{\pi}\int_{0}^{\pi}dq\left[ \sin2q- 2\cos
q\sqrt{1+\sin^{2}q}\right].
\end{eqnarray}
The imaginary part of the integral vanishes, whereas the real part
reduces to
\begin{equation}
\mathcal{I}_{1}=\frac{1}{4}+\frac{1}{2\pi}\int_{-1}^{1}\sqrt{2-x^{2}}=\frac{1}{2}+\frac{1}{2\pi}.
\end{equation}

\section{Integration in \eref{eq:finaldensity} \label{app:I2}}
Let
\begin{eqnarray}
\fl\mathcal{I}_{2}=\int_{0}^{2\pi}d\omega
\gamma(\omega)\times\frac{e^{i\omega
x}}{[z_{+}(\omega)]^{y}}\nonumber\\
\fl\qquad=\int_{0}^{2\pi}d\omega\frac{e^{i\omega (x+1)}-e^{i\omega x}}{\left[\sqrt{\left(
2-\cos\omega\right)^{2}-1}+i\sin \omega\right]\left[
2-\cos\omega+\sqrt{\left( 2-\cos\omega
\right)^{2}-1}\right]^{y}}.
\end{eqnarray}
Performing a change of variable $\omega\rightarrow2q$ and simplifying, the integration
reduces to,
\begin{equation}
\fl\mathcal{I}_{2}=2\int_{0}^{\pi}dq\frac{e^{iq(2x+1)}}{\left( \cos
q-i\sqrt{1+\sin^{2} q} \right) \left( 1+2\sin^{2}q+2\sin
q\sqrt{1+\sin^{2}q} \right)^{y}}.
\end{equation}
Further simplification reduces the integration to
\begin{eqnarray}
\mathcal{I}_{2}=&\int_{0}^{\pi}dq\frac{\cos q\cos \left[ q(2x+1)\right]}{\left(1+2\sin^{2}q+2\sin q\sqrt{1+\sin^{2}q}\right)^{y}}\nonumber\\
&\quad-\int_{0}^{\pi}dq\frac{\sin \left[q(2x+1)\right]\sqrt{1+\sin^{2}q}}{\left(1+2\sin^{2}q+2\sin q\sqrt{1+\sin^{2}q}\right)^{y}}\nonumber\\
&\quad\quad+i\int_{0}^{\pi}dq\frac{\cos q\sin \left[q(2x+1)\right]}{\left(1+2\sin^{2}q+2\sin q\sqrt{1+\sin^{2}q}\right)^{y}}\nonumber\\
&\quad\quad\quad+i\int_{0}^{\pi}dq\frac{\cos \left[q(2x+1)\right]\sqrt{1+\sin^{2}q}}{\left(1+2\sin^{2}q+2\sin q\sqrt{1+\sin^{2}q}\right)^{y}}
\end{eqnarray}
The last two integrands are asymmetric around $q=\pi/2$, and the
integrals vanish. This is also expected as the density difference in
\eref{eq:finaldensity} is a real number. On the other hand the first
two integrands are symmetric around $q=\pi/2$. Then, finally,
\begin{equation}
\mathcal{I}_{2}=2\int_{0}^{\pi/2}dq\frac{\cos q\cos \left[ q(2x+1)\right]-\sin \left[q(2x+1)\right]\sqrt{1+\sin^{2}q}}{\left(1+2\sin^{2}q+2\sin q\sqrt{1+\sin^{2}q}\right)^{y}}
\end{equation}

\section{Asymptotic analysis of the profile in \eref{eq:90}\label{app:excl}}
\subsection{Along $y=0$.}
Let us first consider the profile along the driven lane at $y=0$,
where integral in the expression \eref{eq:90} yields
\begin{eqnarray}
\fl\mathcal{I}_{2}=(1-2\rho) \int_{0}^{\pi/2}dq\frac{\cos\left[
(2x+1)q \right]\cos q}{ 1-2\rho(1-\rho)\cos^{2}q}\nonumber\\
\qquad\qquad\qquad-\int_{0}^{\pi/2}dq\frac{\sqrt{1+\sin^{2}q}}{ 1-2\rho(1-\rho)\cos^{2}q}\sin\left[(2x+1)q\right].
\end{eqnarray}
Using the trigonometric identity $2\cos\left[ (2x+1)q \right]\cos
q=\cos 2xq+\cos\left[ 2(x+1)q \right]$ and a change of variables
$2xq=\eta$, $2(x+1)q=\eta'$
and $(2x+1)q=\xi$ the integral yields
\begin{eqnarray}
\fl\qquad\mathcal{I}_{2}=\frac{ 1-2\rho
}{2x}\int_{0}^{x\pi}d\eta\frac{\cos\eta\cos \left[
\frac{\eta}{2x} \right]}{ 1-2\rho(1-\rho)\cos^{2} \left[
\frac{\eta}{2x} \right]}\nonumber\\
+\frac{ 1-2\rho
}{2(x+1)}\int_{0}^{(x+1)\pi}d\eta'\frac{\cos\eta'\cos \left[
\frac{\eta'}{2(x+)} \right]}{ 1-2\rho(1-\rho)\cos^{2} \left[
\frac{\eta'}{2(x+1)} \right]}\nonumber\\
\qquad\qquad-\int_{0}^{(x+1/2)\pi}d\xi\frac{\sin\xi\sqrt{1+\sin^{2}\frac{\xi}{2x+1}}}{1-2\rho(1-\rho)\cos^{2}\frac{\xi}{2x+1}}.
\end{eqnarray}
For large $x$, the terms involving $x$ in the integrands vary slowly compared
to the rest. Then, in the first two integrals the range of integration
can be divided in intervals of length $\pi$ where within each such
integrals the slowly varying terms can be replaced by their approximate
value within that window. For example, for $\eta\in[n\pi,(n+1)\pi]$
with $n$ being integer,
\begin{equation*}
\frac{\cos(\frac{\eta}{2x})}{1-2\rho(1-\rho)\cos^2(\eta/2x)}\simeq
\frac{\cos(\frac{n\pi}{2x})}{1-2\rho(1-\rho)\cos^2(n\pi/2x)}.
\end{equation*}
Then, it is easy to show that under this approximation the first two
integrals vanish. In the last integral, by dividing the range of
$\xi$ in $0$ to $\pi/2$ and the rest in intervals of $\pi$, it yields
\begin{eqnarray}
\fl\mathcal{I}_{2}\simeq-\frac{1}{1-2\rho(1-\rho)}\int_{0}^{\pi/2}d\xi
\sin\xi\nonumber\\
-\sum_{n=1}^{x}\frac{\sqrt{1+\sin^2\left(
\frac{(2n+1)\pi}{2(2x+1)} \right)}}{1-2\rho(1-\rho)\cos\left(
\frac{(2n+1)\pi}{2(2x+1)} \right)}\int_{(n-1/2)\pi}^{(n+1/2)\pi} d\xi\sin\xi
\end{eqnarray}
The integral in the right most term vanishes for all $n$, wherein the first
integral yields $1$. Then, finally,
\begin{equation}
\mathcal{I}_{2}\simeq-\frac{1}{1-2\rho(1-\rho)}\times\frac{1}{2x+1}.
\end{equation}
Then the asymptotic density profile for large $\vert x\vert$ along the driven lane
\begin{equation}
\phi\left( x,0 \right)-\rho\simeq -\left(
\frac{ \epsilon \phi(-1,0)\left[ 1-\phi(0,0)\right]}{\pi\left[
1-2\rho(1-\rho) \right]} \right)\times \frac{1}{2x+1}.
\end{equation}

\subsection{Along $y\simeq m ~x$.}
Next, we consider the profile on sites along the straight line
$2y=m(2x+1)$. As $y$ is positive for all sites on the lattice, $m\ge0$ for
$x\ge0$ and vise versa. Let us first analyze the case with positive $m$ and
$x$. Along this line the integral in \eref{eq:90} yields
\begin{equation}
\fl\mathcal{I}_{3}=\int_{0}^{\pi/2}dq\frac{(1-2\rho)\cos q(2x+1)\cos
q-\sin q(2x+1)\sqrt{1+\sin^{2} q}}{\left\{
1-2\rho(1-\rho)\cos^{2} q \right\}\left[ 1+2\sin^{2} q+ 2\sin
q\sqrt{1+\sin^{2}q}\right]^{m(x+1/2)}}.
\label{eq:II3}
\end{equation}
The term inside the square bracket in the denominator achieves its
maximum value at $q=0$ and monotonically decreases as $q$ increases
within the range of integration. Then, for large $x$, the leading
contribution in the integral comes from small $q$. In this range,
although $q$ is small, $q(2x+1)$ could be large. By expanding in
terms of $q$ the integrand yields
\begin{eqnarray}
\fl\mathcal{I}_{3}=\int_{0}^{\pi/2}dq\left[\frac{(1-2\rho)\cos
q(2x+1)\times\left\{ 1-\frac{1}{2}q^{2} \right\}-\sin
q(2x+1)\times\left\{ 1+\frac{1}{2}q^{2} \right\}}{\left\{
1-2\rho(1-\rho)\left( 1-q^{2} \right) \right\}\left[
1+2q-\frac{2}{3}q^{3} \right]^{m(x+1/2)}}\right.\nonumber\\
\left.\qquad\qquad\qquad\qquad\qquad\qquad\qquad\qquad\qquad\qquad+\mathcal{O}(q^{3})\right].
\end{eqnarray}
The term in the denominator involving $x$ can be expanded as
$(1+2q-2q^3/3)^{-m(x+1/2)}=\exp[-m(2x+1)q]\{1+m(2x+1)q^3/3+\mathcal{O}(q^4)\}$.
In addition consider an expansion
\begin{equation}
\fl\frac{1}{\left[ 1-2\rho(1-\rho)(1-q^2)
\right]}=\frac{1}{\left[1-2\rho(1-\rho) \right]}\left\{
1-\frac{2\rho(1-\rho)}{1-2\rho(1-\rho)}q^2+\mathcal{O}(q^4)
\right\}
\end{equation}
With these, and keeping only the terms up to order $q^2$, the integral
in terms of a quantity $\xi=(2x+1)q$ yields
\begin{eqnarray}
\fl\mathcal{I}_{3}=\frac{1}{2x+1}\times
\frac{1}{1-2\rho(1-\rho)}\int_{0}^{(2x+1)\frac{\pi}{2}}d\xi \left[
\left( 1-2\rho \right)\cos\xi-\sin\xi \right]e^{-m\xi}\nonumber\\
\fl\qquad+\frac{1}{\left(2x+1\right)^3}\times
\frac{1}{1-2\rho(1-\rho)}\left[\frac{m}{3}\int_{0}^{(2x+1)\frac{\pi}{2}}d\xi \left[
\left( 1-2\rho \right)\cos\xi-\sin\xi \right]\xi^{3}e^{-m\xi}\right.\nonumber\\
\fl\qquad\qquad\left.-\frac{2\rho(1-\rho)}{1-2\rho(1-\rho)}\int_{0}^{(2x+1)\frac{\pi}{2}}d\xi \left[
\left( 1-2\rho \right)\cos\xi-\sin\xi
\right]\xi^{2}e^{-m\xi}\right.\nonumber\\
\left.-\frac{1}{2}\int_{0}^{(2x+1)\frac{\pi}{2}}d\xi \left[
\left( 1-2\rho \right)\cos\xi+\sin\xi
\right]\xi^{2}e^{-m\xi}\right]+\mathcal{O}\left[\frac{1}{(2x+1)^{4}}\right].
\end{eqnarray}
The integrations on the right hand side are easy to perform using method of integration by
parts, whereby the result yields,
\begin{eqnarray}
\fl\left[ 1-2\rho(1-\rho) \right]\mathcal{I}_{3}=\frac{1}{2x+1}\left[
\frac{m(1-2\rho)-1}{m^{2}+1} \right]+\frac{1}{(2x+1)^{3}}\nonumber\\
\qquad\quad\times\left[\frac{m}{3}\left\{6\times\frac{(1-2\rho)+4m-6(1-2\rho)m^{2}-4m^3+(1-2\rho)m^4}{(m^2+1)^{4}}\right\}\right.\nonumber\\
 \qquad\qquad-\frac{2\rho(1-\rho)}{1-2\rho(1-\rho)}\left\{2\times\frac{1-3(1-2\rho)m-3m^2+(1-2\rho)m^3}{(m^2+1)^3}\right\}\nonumber\\
 \qquad\qquad\left.-\frac{1}{2}\left\{
-2\times\frac{1+3(1-2\rho)m-3m^2-(1-2\rho)m^3}{(m^2+1)^3}
\right\}\right]
\end{eqnarray}
Then for $m(1-2\rho)\ne1$, the asymptotic dependence on $x$ is determined by the
leading term which decays as $1/(2x+1)$.

Along the line with slope $m(1-2\rho)=1$ the leading term vanishes, and the sub-leading term determines the profile. By replacing $m$ in terms of
$\rho$, the expression simplifies to
\begin{equation}
\mathcal{I}_{3}\simeq - \frac{\left[ 1+2\rho(1-\rho) \right]\left(
1-2\rho\right)^{4}}{2\times\left[  1-2\rho(1-\rho)
\right]}\times\frac{1}{(2x+1)^{3}}.
\end{equation}

Then from \eref{eq:90}, the asymptotic density profile for $m(1-2\rho)\ne1$,
\begin{equation}
\fl \qquad\phi\left[x,m(x+1/2)\right]-\rho\simeq
\frac{ \epsilon \phi(-1,0)\left[ 1-\phi(0,0)\right]}{\pi\left[1-2\rho(1-\rho)\right]}\times\frac{m(1-2\rho)-1}{m^{2}+1}\times\frac{1}{2x+1}
\end{equation}
and for $m(1-2\rho)=1$,
\begin{equation}
\fl \qquad\phi\left[x,m(x+1/2)\right]-\rho\simeq -
\frac{ \epsilon \phi(-1,0)\left[ 1-\phi(0,0)\right]}{2\pi\left[1-2\rho(1-\rho)\right]}\times\frac{\left[1+2\rho(1-\rho)\right](1-2\rho)^{4}}{\left(2x+1\right)^3}
\end{equation}
The analysis can be easily extended to negative $x$ and $m$. The
expression for the profile remains the same.

\subsection{Along $x=0$.}
The last direction we analyze is along the $x=0$ line perpendicular to
the driven lane, where the integral in \eref{eq:90} yields
\begin{eqnarray}
\fl\mathcal{I}_{4}=\int_{0}^{\pi/2}dq\frac{(1-2\rho) \cos^2 q -\sin
q\sqrt{1+\sin^{2}q}}{ \left\{1-2\rho(1-\rho)\cos^{2}q\right\}\left[ 1+2\sin^{2} q+ 2\sin
q\sqrt{1+\sin^{2}q}\right]^{m(x+1/2)}}.
\end{eqnarray}
Following the same argument as used for the analysis of
\eref{eq:II3} the leading order term in the integral can be
calculated as
\begin{equation}
\mathcal{I}_{4}\simeq\frac{1}{1-2\rho(1-\rho)}\int_{0}^{\pi/2}dq\left[\left(1-2\rho\right)-q+\mathcal{O}(q^2)\right]\exp\left(
-2yq \right).
\end{equation}
Due to the exponential term, the integrand is sharply damped beyond
$q\simeq 1/2y$ and the leading contribution comes from $q$ within this
range. As a result,
\begin{equation}
\left[ 1-2\rho(1-\rho)
\right]\mathcal{I}_{4}\simeq\frac{1-2\rho}{2y}-\frac{1}{8y^2}+\mathcal{O}\left(
y^{-3} \right).
\end{equation}
Then the density profile  \eref{eq:90} along this line yields
\begin{equation}
\fl \qquad\phi\left[0,y\right]-\rho\simeq
\frac{ \epsilon \phi(-1,0)\left[ 1-\phi(0,0)\right]}{\pi\left[1-2\rho(1-\rho)\right]}\times\left[\frac{1-2\rho}{2y}-\frac{1}{8y^2}\right]
\end{equation}

\section{A perturbative solution of the exclusion case. \label{app:higher order terms}}
The mean-field equation in \eref{eq:masterexcl1}-\eref{eq:masterexcl2} is non-linear in $\phi$, and in general difficult to solve. As mentioned earlier, a
method of solving the equations is by perturbative expansion, where
the solution $\phi(x,y)$ is expanded in a series of small parameter
$\epsilon$, as shown in \eref{eq:perturbative}. Perturbation expansions of this type have been studied in recent years to solve coupled non-linear equations
\cite{bender,laurenzi}. The central idea is that, applying this expansion
in the non-linear equation, decomposes it order by order in $\epsilon$
into an infinite sequence of inhomogeneous linear
problems which are all formally solvable. The inhomogeneity in the
equation for the $p$th order depends on the
solution of all the previous $p-1$ orders. In this Appendix, we shall use this perturbative scheme for \eref{eq:masterexcl1}-\eref{eq:masterexcl2} and systematically determine the solution for arbitrary order in $\epsilon$ and extract the large distance density profile from the solution.

Applying the series from \eref{eq:perturbative} into \eref{eq:masterexcl1}-\eref{eq:masterexcl2} and equating the terms of same power in
$\epsilon$, yields
\begin{equation}
\Delta\alpha_{p}(x,y)=0 \textrm{ for \ensuremath{y>0},}
\label{eq:masterexcsl1small}
\end{equation}
and
\begin{eqnarray}
\fl\qquad\alpha _p(x-1,0)+\alpha _p(x,1)-2\alpha _p(x,0)+\rho \left[\alpha
_p(x+1,0)-\alpha _p(x-1,0)\right]\nonumber\\
 \qquad\qquad\qquad\qquad=Q_{p}\left(\delta _{x,0}-\delta
_{x,-1}\right)-F_p\left( x \right).
\label{eq:masterexcsl2small}
\end{eqnarray}
We defined
\begin{equation}
Q_{p}=\alpha _{p-1}(-1,0)-\sum _{q=0}^{p-1} \alpha _q(-1,0)\alpha
_{p-q-1}(0,0),
\label{eq:Qapp}
\end{equation}
with $\alpha_{0}(x,y)=\rho$. The $Q_{p}$ depends only on the lower order solutions $\alpha_{k}$, with $k<p$, at the two adjacent sites of the slow bond.

Note that, using \eref{eq:perturbative} yields,
\begin{equation}
	\sum_{p=1}^{\infty}\epsilon^{p}Q_{p}= \epsilon \phi(-1,0)\left[
	1-\phi(0,0)\right].
	\label{eq:sumQ}
\end{equation}
This identity will be used later in the derivation.

The last quantity $F\left( x \right)$ in \eref{eq:masterexcsl2small}
also depends only on the lower order solutions
\begin{equation}
F_{p}\left( x \right)=\sum _{q=1}^{p-1} \alpha _q(x,0)\left[\alpha
_{p-q}(x+1,0)-\alpha _{p-q}(x-1,0)\right].
\label{eq:F}
\end{equation}

The infinite set of linear equations \eref{eq:masterexcsl1small} and
\eref{eq:masterexcsl2small} can be solved recursively.
The analysis is similar to the solution of
\eref{eq:alpha one master non-zero y}-\eref{eq:alpha one master zero y}. We only present the final solution here.

The solution for the $p$th order in the expansion
\begin{equation}
\fl\alpha_{p}(x,y)=\frac{Q_{p}}{L}\sum_{n=1}^{L-1}\left(
e^{i\omega_{n}}-1
\right)\Gamma_{\rho}(\omega_{n})\frac{e^{i\omega_{n}x}}{\left[z_{+}(\omega_{n})\right]^{y}}+\sum_{n=1}^{L-1}f_{p}(\omega_{n})\Gamma_{\rho}(\omega_{n})\frac{e^{i\omega_{n}x}}{\left[z_{+}(\omega_{n})\right]^{y}}
\label{eq:genexprapp},
\end{equation}
where $\omega_n=2\pi n/L$ with $n=0,1,\cdots,L-1$, and $f_p(\omega_{n})$ is the Fourier transform of $F_p(x)$, defined as
\begin{equation}
f_p(\omega_{n})=\frac{1}{L}\sum_{x=0}^{L-1}F_{p}\left( x \right)e^{-i \omega_n}.
\label{eq:fourierapp}
\end{equation}
The function $\Gamma_{\rho}(\omega_{n})$ is defined in \Eref{eq:Gamma excl}.

The first term in the solution, comes from the term containing the Kronecker
delta functions in \eref{eq:masterexcsl2small}, and the second from
$F_{p}(x)$. For convenience of presentation, let us define
\begin{equation}
a_{p}(\omega_n)=Q_{p}+L\frac{f_{p}(\omega_{n})}{e^{i\omega_{n}}-1}.
\label{eq:aapp}
\end{equation}
Notice $(e^{i\omega_{n}}-1)$ does not vanish in the range of
$1\le n\le L-1$ in the solution. With this the solution \eref{eq:genexprapp} yields,
\begin{equation}
\alpha_{p}(x,y)=\frac{1}{L}\sum_{n=1}^{L-1}\frac{\gamma_{\rho}(\omega_{n})\times
e^{i\omega_{n}x}}{\left[z_{+}(\omega_{n})\right]^{y}}a_{p}(\omega_n),
\label{eq:genexprapp2}
\end{equation}
where
$\gamma_{\rho}(\omega_{n})=(e^{i\omega_{n}}-1)\Gamma_{\rho}(\omega_{n})$.
Notice that for $\rho=0$ the $\gamma_{\rho}(\omega)$ is same as
$\gamma(\omega)$ defined in \eref{eq:gamma} for the independent particles case.

Then, using \eref{eq:perturbative}, yields the expression for the density profile,
\begin{equation}
\phi\left( x,y
\right)-\rho=\frac{1}{L}\sum_{n=1}^{L-1}\frac{\gamma_\rho(\omega_{n})e^{i\omega_{n}x}}{\left[
z_{+}\left( \omega_{n} \right)
\right]^{y}}\times\sum_{p=1}^{\infty}\epsilon^{p}a_{p}\left(
\omega_{n}.
\right)
\label{eq:ph intermediate}
\end{equation}
Using \eref{eq:sumQ} and \eref{eq:aapp}, yields
\begin{equation}	\sum_{p=1}^{\infty}\epsilon^{p}a_{p}\left(\omega_{n}\right)=\epsilon\phi(-1,0)\left[
	1-\phi(0,0)
	\right]+\frac{L}{e^{i\omega_{n}}-1}\sum_{p=1}^{\infty}\epsilon^{p}f_{p}(\omega_{n}).
\end{equation}
Substituting the above in the expression \eref{eq:ph intermediate} yields the density profile,
\begin{eqnarray}
\fl\qquad\qquad\qquad \phi\left( x,y
\right)-\rho=\epsilon\phi(-1,0)\left[
	1-\phi(0,0)
	\right]\frac{1}{L}\sum_{n=1}^{L-1}\frac{\gamma_\rho(\omega_{n})e^{i\omega_{n}x}}{\left[
z_{+}\left( \omega_{n} \right)
\right]^{y}}\nonumber\\
\qquad\qquad\qquad\qquad\qquad\qquad\qquad+\frac{1}{L}\sum_{n=1}^{L-1}\frac{\gamma_\rho(\omega_{n})e^{i\omega_{n}x}}{\left[
z_{+}\left( \omega_{n} \right)
\right]^{y}}\times\Sigma\left( \omega_{n}
\right).
\end{eqnarray}
where
\begin{equation}
	\Sigma\left( \omega_{n}
\right)=\frac{L}{e^{i\omega_{n}}-1}\sum_{p=1}^{\infty}\epsilon^{p}f_{p}(\omega_{n}).
\label{eq:Sigma}
\end{equation}
This is the complete solution of profile, with the $f_p(\omega_{n})$ defined in \eref{eq:fourierapp}. Note, the density difference vanishes as $\epsilon\rightarrow 0$, as
expected.

\subsection{Asymptotic profile}
To study the large distance profile it is simpler to
analyze in the $L\rightarrow \infty$ limit, where
$\omega_n\equiv\omega$ becomes a continuous variable, and the
summation over $n$ can be approximated by an integration. This yields,
\begin{eqnarray}
\fl\qquad\qquad\qquad	\phi\left( x,y
\right)-\rho=\epsilon\phi(-1,0)\left[
	1-\phi(0,0)
	\right]\frac{1}{2\pi}\int_{0}^{2\pi} d\omega \frac{\gamma_\rho(\omega)e^{i\omega x}}{\left[
z_{+}\left( \omega \right)
\right]^{y}}\nonumber\\
\qquad\qquad\qquad\qquad\qquad\qquad+\frac{1}{2\pi}\int_{0}^{2\pi} d\omega \frac{\gamma_\rho(\omega)e^{i\omega x}}{\left[
z_{+}\left( \omega \right)
\right]^{y}}\times\Sigma\left( \omega \right).
\label{eq:densitysuma}
\end{eqnarray}

As shown later in the \ref{app:beta}, the $\Sigma(\omega)$ is analytic around $\omega=0$ and also vanishes
as $\omega\rightarrow 0$. This implies, the $\Sigma(\omega)$ is Taylor expandable with the lowest order being $\omega$.
At large distances, the profile is essentially determined by the small $\omega$
modes. Then clearly, at large distances, the second integral involving $\Sigma(\omega)$ in \eref{eq:densitysuma} is
sub-dominant to the first and the profile is
\begin{equation}
\fl	\qquad\qquad \phi\left( x,y
\right)-\rho\simeq\epsilon\phi(-1,0)\left[
	1-\phi(0,0)
	\right]\frac{1}{2\pi}\int_{0}^{2\pi} d\omega \frac{\gamma_\rho(\omega)e^{i\omega x}}{\left[
z_{+}\left( \omega \right)
\right]^{y}}.
\end{equation}

\subsection{Analyticity of $\Sigma(\omega)$. \label{app:beta}}
We shall show that the $\Sigma(\omega)$ is analytic around $\omega=0$ and has a Taylor expansion in powers of $\omega$. The expression in \eref{eq:Sigma} involves $f_{p}(\omega_{n})$ which can be determined recursively order by order.
However, it is not straightforward to prove the analyticity of $\Sigma(\omega)$ from this expression. We argue in the following steps by expressing $\Sigma(\omega)$ in an alternate form where the $\omega$ dependence is clearer.

A straightforward algebra using \eref{eq:F} and \eref{eq:genexprapp2} yields,
\begin{equation}
	\fl	\qquad \qquad \Sigma(\omega)=\frac{1}{2\pi}\int_{0}^{\omega}d\omega^{\prime}
	 e^{-i\omega^{\prime}}\gamma_\rho(\omega-\omega^{\prime})\gamma_\rho(\omega^{\prime})\sum_{p=1}^{\infty}\epsilon^{p}\sum_{q=1}^{p-1}a_{q}(\omega-\omega^{\prime})a_{p-q}(\omega^{\prime}),
\end{equation}
where we have taken the continuous limit.
The $a_{p}(\omega)$ defined in \eref{eq:aapp} can be determined recursively using
\begin{equation}
a_{1}(\omega_n)=Q_1
\label{eq:a1app}
\end{equation}
and
\begin{equation}
\fl	\qquad \qquad  a_{p}(\omega_{n})=Q_{p}+\frac{1}{2\pi}\int_{0}^{\omega}d\omega^{\prime}
	 e^{-i\omega^{\prime}}\gamma_\rho(\omega-\omega^{\prime})\gamma_\rho(\omega^{\prime})\sum_{q=1}^{p-1}a_{q}(\omega-\omega^{\prime})a_{p-q}(\omega^{\prime}).
	\label{eq:afinal1}
\end{equation}

It is more convenient to express $\Sigma(\omega)$ in terms of the function $\beta_n(\omega)$ defined as
\begin{equation}
\beta_{1}(\omega)=1,\textrm{ ~ ~ ~ ~ ~ ~ ~ and}
\end{equation}
\begin{equation}
\beta_{p}(\omega)=\frac{1}{2\pi}\int_{0}^{\omega}e^{-i\omega}\gamma_\rho(\omega-\omega')\gamma_\rho(\omega)\sum_{q=1}^{p-1}\beta_{q}(\omega)\beta_{p-q}(\omega-\omega'),\textrm{for \ensuremath{p>1}.}
\label{eq:beta p}
\end{equation}
In terms of this, the $a_{p}(\omega)$ in \eref{eq:afinal1} can be expressed as a linear combination of $\beta_p(\omega)$, as follows,
\begin{equation}
a_{1}(\omega)=Q_{1},
\end{equation}
\begin{equation}
a_{2}(\omega)=Q_{2}+Q_{1}^{2}\beta_{2}(\omega),
\end{equation}
\begin{equation}
a_{3}(\omega)=Q_{3}+2Q_{1}Q_{2}\beta_{2}(\omega)+Q_{1}^{3}\beta_{3}(\omega),
\end{equation}
and so on. On the other hand, by definition in \eref{eq:aapp} and  \eref{eq:Sigma},
\begin{equation}
	\sum_{p=1}^{\infty}\epsilon^{p} a_{p}=\sum_{p=1}^{\infty}\epsilon^{p}
	Q_{p}+\Sigma(\omega).
\end{equation}
Then, clearly $\Sigma(\omega)$ can be expressed in a linear combination of
$\beta_{p}$ as
\begin{equation}
\Sigma\left( \omega \right)=A_{2}\beta_{2}(\omega)+A_{3}\beta_{3}(\omega)+\cdots,
\label{eq:SigmaExpand}
\end{equation}
where $A_p$ are constants depending on $Q_p$ and $\epsilon$. In this way the $\omega$ dependence of $\Sigma(\omega)$ is solely in terms of $\beta_p(\omega)$.

In the next step we show that the functions $\beta_{p}(\omega)$ are
analytic around $\omega=0$ and smoothly vanishes as
$\omega\rightarrow0$, for all $p>1$. This can be seen by explicitly writing the
expression of $\gamma_\rho(\omega)$ from \eref{eq:small gamma rho} in the definition of $\beta_{p}(\omega)$ in \eref{eq:beta p}.
This yields,
\begin{eqnarray}
\beta_{p}(\omega)=\frac{e^{i\omega/2}}{2\pi}\int_{0}^{\omega/2}d\omega'&\frac{e^{-i\omega'}}{(1-2\rho)\cos(\frac{\omega'}{2})-i\sqrt{1+\sin^{2}(\frac{\omega'}{2})}}\nonumber\\
&\times\frac{\sum_{q}^{p-1}\beta_{q}(\omega)\beta_{p-q}(\omega-\omega')}{(1-2\rho)\cos(\frac{\omega-\omega'}{2})-i\sqrt{1+\sin^{2}(\frac{\omega-\omega'}{2})}},
\end{eqnarray}
which for small $\omega$ can be approximated as
\begin{equation}
\beta_{p}(\omega)\simeq\frac{e^{i\omega/2}}{2\pi}\int_{0}^{\omega/2}d\omega'\frac{e^{-i\omega'}\sum_{q}^{p-1}\beta_{q}(\omega)\beta_{p-q}(\omega-\omega')}{(1-2\rho-i)^{2}}.
\end{equation}
Then, with $\beta_{1}(\omega)=1$, it is easy to show that
\begin{equation}
	\beta_{p}(\omega)\sim w^{p-1}\qquad\qquad \textrm{as \ensuremath{\omega\rightarrow0}.}
\end{equation}

Using this result in the \Eref{eq:SigmaExpand} yields that $\Sigma(\omega)$ is expandable in positive integer powers of $\omega$ around $\omega=0$, as
\begin{equation}
	\Sigma\left( \omega \right)=C_{2}\omega+C_{3}\omega^{2}+\cdots,
\end{equation}
with $C_{k}$ depending on $Q_{k}$ and $\epsilon$.

\section*{References}
\bibliography{reference}
\bibliographystyle{unsrt}
\end{document}